%% file: main-long.tex
\documentclass[11pt]{article} 

\input{preamble}

\title{Optimal Testing for Properties of Distributions}
\author{Jayadev Acharya\thanks{Supported by a grant from MITEI-Shell program.}\\
EECS, MIT\\
\tt{jayadev@csail.mit.edu}
\and
Constantinos Daskalakis\thanks{Supported by a Sloan Foundation Fellowship, a Microsoft Research Faculty Fellowship and NSF Award CCF-0953960 (CAREER) and CCF-1101491.}\\
EECS, MIT \\
\tt{costis@mit.edu}
\and
Gautam Kamath\thanks{Supported by NSF Award CCF-0953960 (CAREER). Work done in part while the author was at Microsoft Research Cambridge.}\\
EECS, MIT \\
\tt{g@csail.mit.edu}
}

\begin{document}

\maketitle

\input{abstract}

\section{Introduction} \label{sec:introduction}

The quintessential scientific question is whether an unknown object
has some property, i.e. whether a model from a specific class fits  
the object's observed behavior. If the unknown object is
a probability distribution, $\dP$, to which we have sample access,
we are typically asked to distinguish whether $\dP$ belongs to some class 
$\cC$ or whether it is sufficiently far from it.

This question has received tremendous attention in \gnote{the field of} statistics (see, e.g.,~\cite{Fisher25,lehmann2006testing}), where
test statistics for important properties such as the ones we
consider here have been proposed. Nevertheless, the 
emphasis has been on asymptotic
analysis, characterizing \gnote{the} rates of convergence of test statistics under null
hypotheses, as the number of samples tends to infinity. In contrast, 
\gnote{we wish to study the following problem in the small sample regime:}

\vspace{2ex}
\begin{center}
\smallskip \framebox{
\begin{minipage}{13.5cm} $\Pi(\cC,\ve)$: Given a family of distributions $\cC$, some
$\ve>0$, and sample access to an unknown distribution $\dP$ over \gnote{a}
discrete support, how many samples are required to distinguish
between $\dP \in \cC$ versus $\dtv(\dP,\cC)>\ve$?
\end{minipage}
}
\end{center}

\vspace{2ex}
The problem has been studied  intensely in the literature on \gnote{property
testing and sublinear algorithms} \cite{Goldreich98, Fischer01, Rubinfeld06, Ron08,canonne2015survey}, where the emphasis has been on
characterizing the optimal tradeoff between $\dP$'s support size and
the accuracy $\ve$ in the number of samples. Several results have
been obtained, roughly clustering into three groups, where (i) $\cC$
is the class of monotone distributions over $[n]$, or more generally a
poset~\cite{BatuKR04,Bhattacharyya11}; (ii) $\cC$ is the
class of independent, or $k$-wise independent distributions over a
hypergrid~\cite{batu2001testing,alon2007testing}; and
(iii) $\cC$ contains a single-distribution $\dQ$, and the problem
becomes that of testing whether $\dP$ equals $\dQ$ or is far from
it~\cite{batu2001testing,Paninski08, valiant2014automatic}. 

With respect to (iii), \cite{valiant2014automatic} exactly characterizes the number of samples required to test identity to each distribution $\dQ$, providing a single tester matching this bound simultaneously for all $\dQ$. Nevertheless, this tester and its precursors are not applicable to the composite identity testing problem that we consider. If our class $\cC$ were finite, we could test against each element in the class, albeit this would not necessarily be sample optimal. If our class $\cC$ \gnote{were} a continuum, we \gnote{would}  need \emph{tolerant} identity testers, which tend to be more expensive in terms of \gnote{sample complexity} \cite{ValiantV11}, and result in substantially suboptimal testers for the classes we consider. Or we could use approaches related to generalized likelihood ratio test, but their behavior is not well-understood in our regime, and optimizing likelihood over our classes becomes computationally intense.

\paragraph{Our Contributions} In this paper, we obtain sample-optimal and computationally efficient testers for $\Pi(\cC,\ve)$ for the most \gnote{fundamental} shape restrictions to a distribution. 
Our contributions are the following:
\begin{enumerate}
\item
\jnote{For a known distribution $\dQ$ over $[n]$, and given samples from an unknown
distribution $\dP$, we show that distinguishing the cases: $(a)$ whether the $\chi^2$-distance
between $\dP$ and $\dQ$ is at most $\eps^2/2$,  versus $(b)$ the $\ell_1$ distance
between $\dP$ and $\dQ$ is at least $\eps$, requires $\Theta(\sqrt
n/\eps^2)$ samples. 
As a corollary, we provide a simpler argument to
show that identity testing requires $\Theta(\sqrt n/\eps^2)$ samples (previously shown in \cite{valiant2014automatic}). }
\item 
For the class $\cC=\cM$ of monotone distributions over
  $[\absz]^d$ we require an optimal $\Theta\left({\absz^{d/2} \over
      \ve^2}\right)$ number of samples, where prior work requires
  $\Omega\left({\sqrt{n} \log n \over \ve^4}\right)$ samples for $d=1$
  and $\tilde{\Omega}\left(\absz^{d-{1\over 2}} {\rm poly}\left({1 \over
      \ve}\right)\right)$ for $d>1$~\cite{BatuKR04,Bhattacharyya11}. Our
  results improve the exponent of $n$ with respect to $d$, shave all
  logarithmic factors in $n$, and improve the exponent of $\ve$ by at
  least a factor of $2$. 
\begin{enumerate}
\item A useful building block and interesting byproduct of our
  analysis is extending Birg\'e's oblivious decomposition for
  single-dimensional monotone distributions~\cite{Birge87} to monotone
  distributions in $d\ge1$, and to the stronger notion of $\chi^2$-distance. See Section~\ref{sec:hypergrid}. 

\item Moreover, we show that $O(\log^d n)$ samples suffice to learn a monotone
distribution over $[\absz]^d$ in $\chi^2$-distance. See
Lemma~\ref{lem:fin-learn-mon} for the precise statement.
\end{enumerate}
\item
\gnote{For the class $\cC = \prdd$ of product distributions over $[n_1]\times\cdots\times[n_d]$,
our algorithm requires
$O\left(\left((\prod_\ell n_\ell)^{1/2} + \sum_\ell
n_\ell\right)/\eps^2\right)$ samples.
We note that a product distribution is one where all marginals are independent, so this is equivalent to testing if a collection of random variables are all independent.}
\gnote{In the case where
$n_\ell$'s are large, then the first term dominates, and the sample
complexity is $O(\left(\prod_\ell n_\ell\right)^{1/2}/\eps^2)$. 
In particular, when $d$ is a constant and all $n_\ell$'s are equal to $n$, we achieve the optimal sample complexity of $\Theta(n^{d/2}/\eps^2)$.
To the best of our knowledge, this is the first result for $d \geq 3$, and when $d = 2$, this improves the previously known complexity from $O\left(\frac{n}{\eps^6}{\rm
    polylog}(n/\eps)\right)$ \cite{batu2001testing,levi2013testing}, significantly improving the dependence on $\ve$ and shaving all logarithmic factors.
}
\item For the classes $\cC=\cLCD$, $\cC=\cMHR$ and $\cC=\cU$ of log-concave, monotone-hazard-rate and unimodal distributions over $[\absz]$, we require an optimal $\Theta\left({\sqrt{\absz} \over \ve^2}\right)$ number of samples. Our testers for $\cLCD$ and $\cC=\cMHR$ are to our knowledge the first for these classes for the low sample regime we are studying---see~\cite{hall2005testing} and its references for statistics literature on the asymptotic regime. Our tester for $\cU$ improves the dependence of the sample complexity on $\ve$ by at least a factor of $2$ in the exponent, and shaves all logarithmic factors in $n$, compared to testers based on testing monotonicity.
\begin{enumerate}
\item A useful building block and important byproduct of our analysis
  are the first computationally efficient algorithms for properly
  learning log-concave and monotone-hazard-rate distributions, to
  within $\ve$ in total variation distance, from ${\rm
    poly}(1/\ve)$ samples, independent of the domain size
  $\absz$. See Corollaries~\ref{cor:learning LCD}
  and~\ref{cor:learning MHR}. Again, these are the first
  computationally efficient algorithms to our knowledge in the low
  sample regime.~\cite{AcharyaDLS15, ChanDSS13b} provide algorithms for density
  estimation, which are non-proper, i.e. will approximate an unknown
  distribution from these classes with a distribution that does not
  belong to these classes. On the other hand, the statistics
  literature focuses on maximum-likelihood estimation in the
  asymptotic regime---see e.g. \cite{cule2010theoretical} and its
  references. 
\end{enumerate}

\item For all the above classes we obtain matching lower bounds,
  showing that the sample complexity of our testers is optimal with
  respect to $n$, $\ve$ and when applicable $d$. See
  Section~\ref{sec:lower-bounds}. Our lower bounds are based on
  extending Paninski's lower bound for testing uniformity~\cite{Paninski08}. 
\end{enumerate}

At the heart of our tester lies a novel use of the $\chi^2$
statistic. Naturally, the $\chi^2$ and its related $\ell_2$ statistic
have  been used in several of the afore-cited results. We propose a
new use of the $\chi^2$ statistic enabling  our optimal sample
complexity. The essence of our approach is to first draw a small
number of samples (independent of $\absz$ for log-concave and
monotone-hazard-rate distributions and only logarithmic in $\absz$ for
monotone and unimodal distributions) to approximate the unknown
distribution $\dP$ in $\chi^2$ distance. If $\dP \in \cC$, our learner
is required to output a distribution $\dQ$ that is $O(\ve)$-close
to $\cC$ in total variation  and $O(\ve^2)$-close to $\dP$ in
$\chi^2$ distance. Then some analysis reduces our testing problem
to distinguishing the following cases: 
\begin{itemize}
\item 
$p$ and $q$ are $O(\ve^2)$-close in $\chi^2$ distance; this case
corresponds to $\dP \in \cC$. 
\item 
$p$ and $q$ are $\Omega(\ve)$-far in total variation distance; this
case corresponds to $\dtv(\dP,\cC)>\ve$. 
\end{itemize}

\gnote{
We draw a comparison with \emph{robust identity testing}, in which one
must distinguish whether $p$ and $q$ are $c_1\ve$-close or
$c_2\ve$-far in total variation distance, for constants $c_2> c_1 >
0$. 
In \cite{ValiantV11}, Valiant and Valiant show that $\Omega(n/\log n)$
samples are required for this problem -- a nearly-linear sample
complexity, which may be prohibitively large in many settings. 
In comparison, the problem we study tests for $\chi^2$ closeness
rather than total variation closeness: a relaxation of the previous
problem. 
However, our tester demonstrates that this relaxation allows us to
achieve a substantially sublinear complexity of $O(\sqrt{n}/\ve^2)$. 
On the other hand, this relaxation is still tight enough to be useful,
demonstrated by our application in obtaining sample-optimal testers. 
}

\jnote{We note that while the $\chi^2$ statistic for testing
  hypothesis is prevalent in statistics providing optimal error
  exponents in the large-sample regime, to the best of our knowledge,
  in the small-sample regime, \emph{modified-versions} of the $\chi^2$
  statistic have only been recently used for \emph{closeness-testing}
  in~\cite{AcharyaDJOPS12, ChanDVV13} and for testing uniformity of
  monotone distributions in~\cite{AcharyaJOS13}. In
  particular,~\cite{AcharyaDJOPS12} design an unbiased statistic for
  estimating the $\chi^2$ distance between two \emph{unknown}
  distributions.} 

In Section~\ref{sec:testing}, we show that a version of the $\chi^2$
statistic, appropriately excluding certain elements of the support,
is sufficiently well-concentrated to distinguish between the above
cases. Moreover, the sample complexity of our algorithm is optimal for
most classes. 
Our base tester is combined with the afore-mentioned extension of
Birg\'e's decomposition theorem to test monotone distributions in
Section~\ref{sec:monotone} (see Theorem~\ref{thm:monotone-final} 
and Corollary~\ref{cor:high-d}), and is also used to test independence
of distributions in Section~\ref{sec:independence} (see
Theorem~\ref{thm:independence}). 

Naturally, there are several bells and whistles that we need to add to
the above skeleton to accommodate all classes of distributions that we
are considering. For log-concave and monotone-hazard distributions, we
are unable to obtain a cheap (in terms of samples) learner that
$\chi^2$-approximates the unknown distribution $\dP$ throughout its
support. Still, we can identify a subset of the support where the
$\chi^2$-approximation is tight and which captures almost all the
probability mass of $\dP$. We extend our tester to accommodate
excluding subsets of the support from the $\chi^2$-approximation. See
Theorems~\ref{thm:lcd-main} and~\ref{thm:mhr-main} in
Sections~\ref{sec:LCD-main} and \ref{sec:MHR-main}. 

For unimodal distributions, we are even unable to identify a large
enough subset of the support where the $\chi^2$ approximation is
guaranteed to be tight. But we can show that there exists a light
enough piece of the support (in terms of probability mass under $\dP$)
that we can exclude to make the $\chi^2$ approximation tight. Given
that we only use Chebyshev's inequality to prove the concentration of
the test statistic, it would seem that our lack of knowledge of the
piece to exclude would involve a union bound and a corresponding
increase in the required number of samples. We avoid this through a
careful application of Kolmogorov's max inequality in our setting. See
Theorem~\ref{thm:unimodality} of Section~\ref{sec:unimodal-main}. 

\paragraph{Related Work.} \jnote{For the problems that we study in
  thie paper, we have provided the 
related works in the previous section along with our contributions.} 
We cannot do justice to the role of shape
restrictions of probability distributions in probabilistic modeling
and testing. 
It suffices to say that the classes of distributions that
we study are fundamental, motivating extensive literature on their
learning and testing~\cite{BBBB:72}. In the recent times, there has
been work on shape restricted statistics, pioneered by Jon Wellner,
and others.~\cite{JW:09,BW10sn} study estimation of monotone and $k-$
monotone densities, and ~\cite{BJR11,SumardW14} study estimation of
log-concave distributions.

As we have mentioned, statistics has focused on the asymptotic
regime as the number of samples tends to infinity. Instead we are
considering the low sample regime and  are more stringent about the
behavior of our testers, requiring $2$-sided guarantees. We want to
accept if the unknown distribution is in our class of interest, and
also reject if it is far from the class. For this problem, as
discussed above, there are few results when $\cC$ is a whole class of
distributions. Closer related to our paper is the line of
papers~\cite{BatuKR04, ACS10, Bhattacharyya11} for monotonicity testing, albeit these 
papers have sub-optimal sample complexity as discussed above.
\gnote{Testing independence of random variables has a long history in
  statisics~\cite{rao1981analysis, agresti2011categorical}. The theoretical
  computer science community has also considered 
  the problem of testing independence of two random
  variables~\cite{batu2001testing, levi2013testing}.  
While our results sharpen the case where the variables are over domains of equal size,
they demonstrate an interesting asymmetric upper bound when this is not the case.} 
More recently, Acharya and Daskalakis provide optimal testers for the
family of Poisson Binomial Distributions~\cite{AcharyaD15}. 

\gnote{
Finally,
contemporaneous work of Canonne et al~\cite{CanonneDGR15a,CanonneDGR15b} provides
a generic algorithm and lower bounds for the single-dimensional families of distributions considered here.
We note that their algorithm has a sample complexity which is suboptimal in both $n$ and $\ve$, while our algorithms are optimal.
Their algorithm also extends to mixtures of these classes, though some of these extensions are not computationally efficient.
They also provide a framework for proving lower bounds, giving the optimal bounds for many classes when $\ve$ is sufficiently large with respect to $1/n$. 
In comparison, we provide these lower bounds unconditionally by modifying Paninski's construction \cite{Paninski08} to suit the classes we consider.
}

\input{preliminaries}
\input{overview}

\input{testing}

\input{monotone}
\input{unimodal-main}

\input{independence}

\input{otherclasses}

\input{lower-bounds}

\section*{Acknowledgements}
The authors thank Cl\'ement Canonne and Jerry Li; the former for several useful comments and suggestions on previous drafts of this work, and both for helpful discussions and thoughts regarding independence testing.
\bibliographystyle{alpha}
\bibliography{biblio}
\appendix
\input{testing-appendix}

\input{monotone-appendix}

\input{unimodal-appendix}
\input{independence-appendix}
\input{lcd-appendix}
\input{mhr-appendix}
\input{lb-appendix}‏

\end{document}

%% file: preamble.tex
\usepackage{ifthen}
\usepackage{mdwlist}
\usepackage{amsmath,amssymb,amsfonts,amsthm}
\usepackage{bm}
\usepackage{mathtools}
\usepackage{url}
\usepackage{hyperref}
\usepackage{enumitem}
\usepackage{graphicx}
\usepackage{xspace}
\usepackage{verbatim}
\usepackage{algorithm}
\usepackage{algpseudocode}
\usepackage[margin=1in]{geometry}
\usepackage[usenames,dvipsnames,svgnames,table]{xcolor}

\usepackage[colorinlistoftodos,textsize=scriptsize]{todonotes}

\def\d{\delta}

\def\eps{\varepsilon}
\def\ve{\varepsilon}

\def\g{\gamma}

\def\l{\lambda}

\def\R{\Rho}
\def\s{\sigma}

\def\to{\rightarrow}

\def\R{{\bf  R}}

\def\cX{{\cal X}}

\newcommand{\prob}[2][]{\text{\bf Pr}\ifthenelse{\not\equal{}{#1}}{_{#1}}{}\!\left[#2\right]}
\newcommand{\expect}[2][]{\text{\bf E}\ifthenelse{\not\equal{}{#1}}{_{#1}}{}\!\left[#2\right]}

\newcommand{\dtv}{d_{\mathrm {TV}}}
\newcommand{\dk}{d_{\mathrm K}}

\newcommand{\ed}{\stackrel{\mathrm{def}}{=}}

\newtheorem{theorem}{Theorem}

\newtheorem{proposition}{Proposition}

\newtheorem{remark}{Remark}

\newtheorem{lemma}{Lemma}
\newtheorem{claim}{Claim}
\newtheorem{corollary}{Corollary}

\newtheorem{definition}{Definition}

\newcommand{\ignore}[1]{}
\providecommand{\poly}{\operatorname*{poly}}

\newenvironment{prevproof}[2]{\noindent {\em {Proof of {#1}~\ref{#2}:}}}{$\hfill\qed$\vskip \belowdisplayskip}

\newcommand{\bg}[1]{\medskip\noindent{\bf #1}}

\definecolor{Red}{rgb}{1,0,0}
\definecolor{Blue}{rgb}{0,0,1}

\newcommand{\costasnote}[1]{#1}

\newcommand{\gnote}[1]{#1}

\newcommand{\jnote}[1]{{#1}}

\newcommand{\oldbound}[1]{{}}

\newcommand{\expectation}[1]{\mathbb{E}\left[#1\right]}
\newcommand{\probof}[1]{{\rm Pr}\left(#1\right)}
\newcommand{\poi}{\rm Poisson}

\newcommand{\E}[1]{\expectation{#1}}
\newcommand{\Var}[1]{\operatorname{Var}\left[#1\right]}
\newcommand{\nsmp}{m}
\newcommand{\absz}{n}
\newcommand{\compl}[1]{\bar{#1}}
\newcommand{\cA}{\mathcal{A}}
\newcommand{\cC}{\mathcal{C}}

\newcommand{\accept}{\textsc{Accept}\xspace}
\newcommand{\reject}{\textsc{Reject}\xspace}
\newcommand{\dP}{p}
\newcommand{\dQ}{q}
\newcommand{\dPi}{\dP_i}
\newcommand{\dPj}{\dP_j}
\newcommand{\dQi}{\dQ_i}
\newcommand{\dQipo}{\dQ_{i+1}}
\newcommand{\dPipo}{\dP_{i+1}}
\newcommand{\prdd}{\Pi_d}

\newcommand{\dQbar}{\bar{\dQ}}
\newcommand{\dPbar}{\bar{\dP}}
\newcommand{\dPbari}{\bar{\dP}_i}
\newcommand{\dPbi}{{\dP}_{\bi}}
\newcommand{\dQbi}{{\dQ}_{\bi}}

\newcommand{\dPbarbi}{{\dPbar}_{\bi}}
\newcommand{\dPbj}{{\dP}_{\bj}}

\newcommand{\bj}{{\bf j}}
\newcommand{\bi}{{\bf i}}
\newcommand{\be}{{\bf e}}

\newcommand{\largeset}{\mathcal{S}_{\rm large}}

\newcommand{\cLCD}{\mathcal{LCD}_n}
\newcommand{\cM}{\mathcal{M}_n^d}
\newcommand{\cU}{\mathcal{U}_n}
\newcommand{\cMHR}{\mathcal{MHR}_n}

\newcommand{\cQ}{\mathcal{Q}}

\newcommand{\chisqpq}[2]{\chi^2(#1,#2)}

%% file: abstract.tex
\begin{abstract}
Given samples from an unknown  distribution $\dP$, is 
it possible to distinguish whether $\dP$ belongs 
to some class of distributions $\cC$ versus $\dP$ being far from 
every distribution in $\cC$? This fundamental question has received
tremendous attention in statistics, focusing primarily on
asymptotic analysis, and more recently in information theory and theoretical computer
science, where the emphasis has been on small sample size and computational
complexity. Nevertheless, even for basic properties of
distributions such as monotonicity, log-concavity, unimodality,
independence, and monotone-hazard rate, the 
optimal sample complexity is unknown.

We provide a general approach via which we obtain sample-optimal and
computationally efficient testers for all these distribution
families. 
\gnote{At the core of our approach is an algorithm which solves the following problem:
Given samples
from an unknown distribution $\dP$, and a known distribution $\dQ$, 
are $\dP$ and $\dQ$ close in $\chi^2$-distance, or far in total variation distance?
}
\gnote{With this tool in place, we develop a general testing framework which leads to the following results:}
\begin{itemize}
\item
\jnote{Testing identity to any distribution over $[\absz]$ 
requires $\Theta(\sqrt{\absz}/\ve^2)$ samples. This is optimal for the
uniform distribution. This gives an alternate argument
for the minimax sample complexity of testing identity (proved
in~\cite{valiant2014automatic}).}
\item 
For all $d \geq 1$ and $n$ sufficiently large, testing whether a discrete distribution
over $[\absz]^d$ is monotone requires an optimal $\Theta(\absz^{d/2}/\ve^2)$ samples.
The single-dimensional version of our theorem improves a long line of 
research starting with~\cite{BatuKR04}, where the \gnote{previous} best
tester required $\Omega(\sqrt{n} \log (n) / \ve^4)$ samples, while the high-dimensional 
version improves~\cite{Bhattacharyya11}, which requires
$\tilde{\Omega}(\absz^{d-{1\over 2}} {\rm poly}({1 \over \ve}))$
samples. 
\item
\jnote{For all $d\ge1$, testing whether a collection of random variables over
$[n_1]\times\cdots\times[n_d]$ are independent requires \gnote{$O\left(\left((\prod_l n_l)^{1/2} + \sum_l
n_l\right)/\eps^2\right)$} samples. A lower bound of
$\Omega\left((\prod_l n_l)^{1/2}/\eps^2\right)$ samples is also proved. This
result extends the known results for testing independence to more than
two random variables. For the special case of $d=2$, when $n_1=n_2=n$,
this improves the results of~\cite{batu2001testing} to the optimal
$\Theta(n/\eps^2)$ \gnote{sample complexity}.}  
\item 
Testing whether a discrete distribution over $[\absz]$ is log-concave
requires an optimal $\Theta(\sqrt{\absz}/\ve^2)$ samples. The
same is true for testing whether a distribution has \gnote{a} monotone hazard rate, and testing
whether it is unimodal.
\end{itemize}
The optimality of \gnote{our} testers is established by providing matching lower
bounds with respect to both $n$ and $\ve$. Finally, a necessary building block 
for our testers and an important byproduct of our work are the
first known computationally efficient proper learners for discrete
log-concave and monotone hazard rate distributions. 

\end{abstract}

%% file: preliminaries.tex
\section{Preliminaries}


We use the following probability distances in our paper.
\begin{definition}
The \emph{total variation distance} between distributions
$p$ and $q$ is defined as 
$$\dtv(p,q) \ed \sup_{A} |p(A) - q(A)| = \frac12\|p - q\|_1.$$
\end{definition}
For a subset of the domain, the total variation distance is defined as
half of the $\ell_1$ distance restricted to the subset.  

\begin{definition} \label{def:chisq}
The \emph{$\chi^2$-distance} between $p$ and $q$ over $[n]$ is defined by
$$  \chi^2(p,q) \ed \sum_{i \in [n]}\frac{(p_i-q_i)^2}{q_i} = \left[\sum_{i \in [n]}\frac{p_i^2}{q_i}\right]-1.$$
\end{definition}

\begin{definition}
  The \emph{Kolmogorov distance} between two probability measures
  $\dP$ and $\dQ$ over an ordered set ($e.g.$, $\R$) with cumulative
  density functions (CDF) $F_p$ and $F_q$ is defined as
  $$\dk(p,q) \ed \sup_{x \in \mathbb{R}} |F_p(x) - F_q(x)|.$$
\end{definition}

Our paper is primarily concerned with testing against classes of distributions, defined formally as follows:
\begin{definition}
Given $\ve \in (0,1]$ and sample access to a distribution $p$, an algorithm is said to \emph{test} a class $\cC$ if it has the following guarantees:
\begin{itemize}
\item If $p \in \cC$, the algorithm outputs \accept with probability at least $2/3$;
\item If $\dtv(p,\cC) \geq \ve$, the algorithm outputs \reject with probability at least $2/3$.
\end{itemize}
\end{definition}

The Dvoretzky-Kiefer-Wolfowitz (DKW) inequality gives a generic algorithm for learning any distribution with respect to the Kolmogorov distance~\cite{DvoretzkyKW56}. 
\begin{lemma}{(See~\cite{DvoretzkyKW56},\cite{Massart90})}
 \label{lem:dkw}
  Suppose we have $n$ \textit{i.i.d.} samples $X_1, \dots X_n$ from a distribution with CDF $F$.
  Let $F_n(x) \ed \frac{1}{n}\sum_{i=1}^n \mathbf{1}_{\{X_i \leq x\}}$ be the empirical CDF.
  Then $\Pr[\dk(F,F_n) \geq \ve] \leq 2e^{-2n\ve^2}$.
  In particular, if $n = \Omega((1/\ve^2) \cdot \log(1/\d))$, then $\Pr[\dk(F,F_n) \geq \ve] \leq \d$.
\end{lemma}

We note the following useful relationships between these distances~\cite{GibbsS02}:
\begin{proposition}
\label{prop:distance-relations}
  $\dk(p,q)^2 \leq \dtv(p,q)^2 \leq \frac14 \chi^2(p,q)$.
\end{proposition}

In this paper, we will consider the following classes of distributions:
\begin{itemize}
\item Monotone distributions over $[n]^d$ (denoted by $\cM$), for which $i \lesssim j$ implies $f_i \geq f_j$\footnote{This definition describes monotone non-increasing distributions. By symmetry, identical results hold for monotone non-decreasing distributions.};
\item Unimodal distributions over $[n]$ (denoted by $\cU$), for which there exists an $i^*$ such that $f_i$ is non-decreasing for $i \leq i^*$ and non-increasing for $i \geq i^*$;
\item Log-concave distributions over $[n]$ (denoted by $\cLCD$), the
  sub-class of unimodal distributions for which $f_{i-1}f_{i+1} \leq f_i^2$; 
\item Monotone hazard rate (MHR) distributions over $[n]$ (denoted by $\cMHR$), for which $i < j$ implies $\frac{f_i}{1 - F_i} \leq \frac{f_j}{1 - F_j}$.
\end{itemize}


\begin{definition}
An \emph{$\eta$-effective support} of a distribution $p$ is any set $S$ such that $p(S) \geq 1 - \eta$. 
\end{definition}

The \emph{flattening} of a function $f$ over a subset $S$ is the function $\bar{f}$ such that
$\bar{f}_i= p(S)/|S|$. 

\begin{definition}
\label{def:flattening}
Let $p$ be a distribution, and support $I_1, \ldots$ is a partition of
the domain. The flattening of $p$ with respect to $I_1,
\ldots$ is the distribution $\bar p$ which is the flattening of $p$
over the intervals $I_1, \ldots$. 
\end{definition}

\paragraph{Poisson Sampling}
Throughout this paper, we  use the standard Poissonization
approach.
Instead of drawing exactly $m$ samples from a distribution $p$, we first draw $m' \sim \poi(m)$, and then draw $m'$ samples from $p$.
As a result, the number of times different elements in the support of $p$ occur in the sample become independent, giving much simpler analyses.
In particular, the number of times we will observe domain element $i$ will be distributed as $\poi(m\dPi)$, independently for each $i$.
Since $\poi(m)$ is tightly concentrated around $m$, this additional flexibility comes only at a sub-constant cost in the sample complexity
with an inversely exponential in $m$, additive increase in the error probability. 

%% file: overview.tex
\section{Overview}
Our algorithm for testing a distribution $p$ can be decomposed into three steps.

\paragraph{Near-proper learning in $\chi^2$-distance.}
Our first step requires a learning algorithm with very specific guarantees.
In proper learning, we are given sample access to a distribution $p \in \cC$, where $\cC$ is some class of distributions, and we wish to output $q \in \cC$ such that $p$ and $q$ are close in total variation distance.
In our setting, given sample access to $p \in \cC$, we wish to output $q$ such that $q$ is \emph{close} to $\cC$ in total variation distance, and $p$ and $q$ are close in $\chi^2$-distance on an effective support\footnote{We also require the algorithm to output a description of an effective support for which this property holds.
This requirement can be slightly relaxed, as we show in our results for testing unimodality.} of $p$.
From an information theoretic standpoint, this problem is harder than proper learning, since $\chi^2$-distance is more restrictive than total variation distance.
Nonetheless, this problem can be shown to have comparable sample complexity to proper learning for the structured classes we consider in this paper.

\paragraph{Computation of distance to class.}
The next step is to see if the hypothesis $q$ is close to the class $\cC$ or not.
Since we have an explicit description of $q$, this step requires no further samples from $p$, i.e. it is purely computational.
If we find that $q$ is far from the class $\cC$, then it must be that $p \not \in \cC$, as otherwise the guarantees from the previous step would imply that $q$ is close to $\cC$. Thus, if it is not, we can terminate the algorithm at this point.

\paragraph{$\chi^2$-testing.}
At this point, the previous two steps guarantee that our distribution $q$ is such that:
\begin{itemize}
\item If $p \in \cC$, then $p$ and $q$ are close in $\chi^2$ distance on a (known) effective support of $p$;
\item If $\dtv(p,\cC) \geq \ve$, then $p$ and $q$ are far in total variation distance.
\end{itemize}
We can distinguish between these two cases using $O(\sqrt{n}/\ve^2)$ samples with a simple statistical $\chi^2$-test, that we describe in Section~\ref{sec:testing}.

\smallskip Using the above three-step approach, our tester, as described in the next section, can directly test monotonicity, log-concavity, and monotone hazard rate. With an extra trick, using Kolmogorov's max inequality, it can also test unimodality.

%% file: testing.tex
\section{A Robust $\chi^2$-$\ell_1$ Identity Test} \label{sec:testing}

{
Our main result in the Section is Theorem~\ref{thm:chisq-test}. 
As an immediate corollary, we obtain the following result on testing
whether an unknown distribution is close in $\chi^2$ or far in
$\ell_1$ distance to a known distribution. In
particular, we show the following:
\begin{theorem}
\label{thm:rob-iden}
For a known distribution $\dQ$, there exists an algorithm with sample
complexity
\[
O(\sqrt n/\eps^2)
\]
distinguishes between the cases
\begin{itemize}
\item 
$\chi^2(\dP,\dQ)<\eps^2/10$\ \ \ \ \emph{versus}
\item
$\|\dP-\dQ\|>\eps^2$.
\end{itemize}
with probability at least $5/6$.
\end{theorem}
This theorem follows from our main result of this section, stated
next, slightly more generally for classes of distributions. 
}
\begin{theorem}
\label{thm:chisq-test}
Suppose we are given $\ve \in (0,1]$, a class of probability distributions $\cC$, sample access to a distribution $p$ over $[n]$, and an explicit description of a distribution $q$ with the following properties:
\begin{enumerate}[label=\textbf{Property \arabic*.},ref=Property \arabic*,align=left]
\item $\dtv(q,\cC) \leq \frac{\ve}{2}$.\label{prp:q-tv}
\item If $p \in \cC$, then $\chi^2(p,q) \leq \frac{\ve^2}{500}$. \label{prp:in-chisq}
\end{enumerate}
Then there exists an algorithm with the following guarantees:
\begin{itemize}
\item If $p \in \cC$, the algorithm outputs \accept with probability at least $2/3$;
\item If $\dtv(p,\cC) \geq \ve$, the algorithm outputs \reject with probability at least $2/3$.
\end{itemize}
The time and sample complexity of this algorithm are $O\left(\frac{\sqrt{n}}{\ve^2}\right)$.
\end{theorem}

\begin{remark}
\label{rmk}
As stated in Theorem~\ref{thm:chisq-test}, \ref{prp:in-chisq} requires that $q$ is $O(\ve^2)$-close in $\chi^2$-distance to $p$ over its entire domain.
For the class of monotone distributions, we are able to efficiently obtain such a $q$, which immediately implies sample-optimal learning algorithms for this class.
However, for some classes, we cannot learn a $q$ with such strong guarantees, and we must consider modifications to our base testing algorithm.

For example, for log-concave and monotone hazard rate distributions, we can obtain a distribution $q$ and a set $S$ with the following guarantees:
\begin{itemize}
\item If $p \in \cC$, then $\chi^2(p_S,q_S) \leq O(\ve^2)$ and $p(S) \geq 1 - O(\ve)$;
\item If $\dtv(p,\cC) \geq \ve$, then $\dtv(p,q) \geq \ve/2$.
\end{itemize}
In this scenario, the tester will simply pretend the support of $p$ and $q$ is $S$, ignoring any samples and support elements in $[n] \setminus S$.
Analysis of this tester is extremely similar to what we present below.
In particular, we can still show that the statistic $Z$ will be separated in the two cases.
When $p \in \cC$, excluding $[n] \setminus S$ will only reduce $Z$.
On the other hand, when $\dtv(p,\cC) \geq \ve$, since $p(S) \geq 1 - O(\ve)$, $p$ and $q$ must still be far on the remaining support, and we can show that $Z$ is still sufficiently large.
Therefore, a small modification allows us to handle this case with the same sample complexity of $O(\sqrt{n}/\ve^2)$.

A further modification can handle even weaker learning guarantees.
We could handle the previous case because the tester ``knows what we don't know'' -- it can explicitly ignore the support over which we do not have a $\chi^2$-closeness guarantee. 
A more difficult case is when there may be a low measure interval hidden in our effective support, over which $p$ and $q$ have a large $\chi^2$-distance.
While we may have insufficient samples to reliably identify this interval, it may still have a large effect on our statistic.
A naive solution would be to consider a tester which tries all possible ``guesses'' for this ``bad'' interval, but a union bound would incur an extra logarithmic factor in the sample complexity.
We manage to avoid this cost through a careful analysis involving Kolmogorov's max inequality, maintaining the $O(\sqrt{n}/\ve^2)$ sample complexity even in this more difficult case. 

Being more precise, we can handle cases where we can obtain a distribution $q$ and a set of intervals $S = \{I_1,\dots, I_b\}$ with the following guarantees:
\begin{itemize}
\item If $p \in \cC$, then $p(S) \geq 1 - O(\ve)$, $p(I_j) = \Theta(p(S)/b)$ for all $j \in [b]$, and there exists a set $T \subseteq [b]$ such that $|T| \geq b - t$ (for $t = O(1)$) and $\chi^2(p_R,q_R) \leq O(\ve^2)$, where $R = \cup_T I_j$;
\item If $\dtv(p,\cC) \geq \ve$, then $\dtv(p,q) \geq \ve/2$.
\end{itemize}
This allows us to additionally test against the class of unimodal distributions.

The tester requires that an effective support is divided into several intervals of roughly equal measure.
It computes our statistic over each of these intervals, and we let our statistic $Z$ be the sum of all but the largest $t$ of these values.
In the case when $p \in \cC$, $Z$ will only become smaller by performing this operation.
We use Kolmogorov's maximal inequality to show that $Z$ remains large when $\dtv(p,\cC) \geq \ve$. 
More details on this tester are provided in Section~\ref{sec:unimodal-appendix}.
\end{remark}


\begin{algorithm}[h]
 \caption{Chi-squared testing algorithm}\label{alg:testing}
\begin{algorithmic}[1]
\State \textbf{Input:} $\ve$; an explicit distribution $q$; (Poisson) $m$ samples from a distribution $p$, where $N_i$ denotes the number of occurrences of the $i$th domain element.
\State $\mathcal{A} \leftarrow \{i:q_i \geq \ve/50n\}$
\State $Z \leftarrow \sum_{i \in \mathcal{A}} \frac{(N_i - mq_i)^2 - N_i}{mq_i}$
\If {$Z \leq m\ve^2/10$}
\State \Return \accept
\Else 
\State \Return \reject
\EndIf 
\end{algorithmic}
\end{algorithm}

\begin{prevproof}{Theorem}{thm:chisq-test}
Theorem \ref{thm:chisq-test} is proven by analyzing Algorithm \ref{alg:testing}.
As shown in Section~\ref{sec:chisq-moments}, $Z$ has the following mean and variance:
\begin{equation}
\E{Z} = m \cdot \sum_{i \in \mathcal{A}} \frac{(p_i - q_i)^2}{q_i} = m \cdot \chi^2(p_\cA,q_\cA)  \label{eqn:mean}
\end{equation}
\begin{equation}
\Var{Z} = \sum_{i \in \mathcal{A}}\left[2 \frac{p_i^2}{q_i^2}+4m\cdot\frac{p_i\cdot(p_i-q_i)^2}{q_i^2}\right]  \label{eqn:variance}
\end{equation}
where by $p_\cA$ and $q_\cA$ we denote respectively the vectors $p$ and $q$ restricted to the coordinates in $\cA$, and we slightly abuse notation when we write $\chi^2(p_\cA,q_\cA)$, as these do not then correspond to probability distributions. 

Lemma~\ref{lem:means} demonstrates the separation in the means of the statistic $Z$ in the
two cases of interest, $i.e.,$ $p \in \cC$ versus $\dtv(p,\cC) \geq
\ve$, and Lemma~\ref{lem:vars} shows the separation in the variances in the
two cases. These two results are proved in
Section~\ref{sec:chisq-analysis}. 

\begin{lemma}
\label{lem:means}
If $p \in \cC$, then $\E{Z} \leq \frac{1}{500}m\ve^2$.
If $\dtv(p,\cC) \geq \ve$, then $\E{Z} \geq \frac{1}{5}m\ve^2$.
\end{lemma}

\begin{lemma}
\label{lem:vars}
If $p \in \cC$, then $\Var{Z} \leq \frac{1}{500000}m^2\ve^4$.
If $\dtv(p,\cC) \geq \ve$, then $\Var{Z} \leq \frac{1}{100}E[Z]^2$.
\end{lemma}

Assuming Lemmas~\ref{lem:means} and~\ref{lem:vars},
Theorem~\ref{thm:chisq-test} is now a simple application of
Chebyshev's inequality. 

When $p \in \cC$, we have that
$$\E{Z} + \sqrt{3}\Var{Z}^{1/2} \leq \left(\frac{1}{500} + \sqrt{3}\left(\frac{1}{500000}\right)^{1/2}\right)m\ve^2 \leq \frac{1}{200}m\ve^2.$$
Thus, Chebyshev's inequality gives
$$\Pr\left[Z \geq m\ve^2/10\right] \leq \Pr\left[Z \geq m\ve^2/200\right] \leq \Pr\left[Z - \E{Z} \geq \sqrt{3}\Var{Z}^{1/2}\right] \leq \frac13.$$

The case for $\dtv(p,\cC) \geq \ve$ is similar.
Here,
$$\E{Z} - \sqrt{3}\Var{Z}^{1/2} \geq \left(1 - \sqrt{3}\left(\frac{1}{100}\right)^{1/2}\right)E[Z] \geq 3m\ve^2/20.$$
Therefore,
\[\Pr\left[Z \leq m\ve^2/10\right] \leq \Pr\left[Z \leq 3m\ve^2/20\right] \leq \Pr\left[Z - \E{Z} \leq - \sqrt{3}\Var{Z}^{1/2}\right] \leq \frac13.\qedhere\qedhere\]
\end{prevproof}

%% file: monotone.tex
\section{Testing Monotonicity}
\label{sec:monotone}
As an application of our testing framework, we will demonstrate how to
test for monotonicity. Let $d\geq1$, and $\bi=(i_1, \ldots,
i_d),\bj=(j_1, \ldots, j_d)\in[n]^d$. We say
$\bi\succcurlyeq\bj$ if $i_l>j_l$ for $l=1,\ldots,d$. 
\begin{definition}
A distribution $p$ over $[n]^d$
is monotone (decreasing) if for all $\bi\succcurlyeq\bj$, $p_{\bi}\le
p_{\bj}$. 
\end{definition}
Our main result of this section is as follows:
\begin{theorem}
\label{thm:monotone-final}
For any $d \geq 1$, there exists an algorithm for testing monotonicity over $[n]^d$ with sample complexity
$$O\left(\frac{\absz^{d/2}}{\eps^2}+\left(\frac{d\log \absz}{\eps^2}\right)^d\cdot \frac1{\eps^2}\right)$$
and time complexity $O\left(\frac{n^{d/2}}{\ve^2} + \poly(\log n, 1/\ve)^d \right)$. 
\end{theorem}

In particular, this implies the following optimal algorithms for
monotonicity testing for all $d \geq 1$: 
\begin{corollary}
\label{cor:high-d}
Fix any $d \geq 1$, and suppose $\eps>\frac{\sqrt{d\log n}}{n^{1/4}}$. 
Then there exists an algorithm for testing monotonicity over $[n]^d$
with sample complexity $O\left(n^{d/2}/\ve^2\right)$. 
\end{corollary}

Our analysis starts with a structural lemma about monotone distributions.
In \cite{Birge87}, Birg\'e showed that any monotone distribution $p$
over $[n]$ can be \emph{obliviously} decomposed into $O(\log(n)/\ve)$
intervals, such that the flattening $\bar p$ (recall
Definition~\ref{def:flattening}) of $p$ over these intervals is 
$\ve$-close to $p$ in total variation distance. 
\cite{AcharyaJOS14a} extend this result, giving a bound between the
$\chi^2$-distance of $p$ and $\bar p$. 
We strengthen these results by extending them to monotone distributions over $[n]^d$.
In particular, we partition the domain $[n]^d$ of $p$ into $O(
(d\log(n)/\ve^2)^d)$ rectangles, and compare it with $\bar p$, the
flattening over these rectangles. 

\begin{lemma}
\label{lem:mon-str}
Let $d\ge1$. There is an oblivious decomposition of $[n]^d$ into
$O((d\log(n)/\ve^2)^d)$ rectangles such that for any monotone
distribution $p$ over $[n]^d$, its flattening $\bar p$ over these
rectangles satisfy $\chi^2(p,\bar p) \leq \ve^2$.  
\end{lemma}

This effectively reduces the support size to logarithmic in $n$.
At this point, we can apply the Laplace estimator (along the lines
of~\cite{KamathOPS15}) and learn a $q$ such
that if $p$ was monotone, then $q$ will be $O(\ve^2)$-close in
$\chi^2$-distance. 

\begin{lemma}
\label{lem:fin-learn-mon}
Let $d\ge1$, and $\dP$ be a monotone distribution over $[\absz]^d$. 
There is an algorithm which outputs a distribution $\dQ$ such that $\expectation{\chi^2(\dP,\dQ)}\le \frac{\eps^2}{500}$.
The time and sample complexity are both $O((d\log (\absz)/\eps^2)^d/\eps^2)$.
\end{lemma}

The final step before we apply our $\chi^2$-tester is to compute the
distance between $q$ and $\cM$. 
This subroutine is similar to the one introduced by~\cite{BatuKR04}. 
The key idea is to write a linear program, which searches for any distribution $f$ which is close to $q$ in total variation distance.
We note that the desired properties of $f$ (i.e., monotonicity, normalization, and $\ve$-closeness to $q$) are easy to enforce as linear constraints.
If we find that such an $f$ exists, we will apply our $\chi^2$-test to $q$.
If not, we output $\reject$, as this is sufficient evidence to conclude that $p \not \in \cM$.
Note that the linear program operates over the oblivious decomposition
used in our structural result, so the complexity is polynomial in
$(d\log(n)/\ve)^d$, rather than the naive $n^d$. 

At this point, we have precisely the guarantees needed to apply Theorem~\ref{thm:chisq-test}, directly implying Theorem~\ref{thm:monotone-final}.
Proof of the lemmas in this section are provided in Section~\ref{sec:monotone-appendix}.

%% file: unimodal-main.tex
We note that the class of monotone distributions is the simplest of the classes we consider.
We now consider testing for log-concavity, monotone hazard rate, and unimodality, all of which are much more challenging to test.
In particular, these classes require a more sophisticated structural understanding, more complex proper $\chi^2$-learning algorithms, and non-trivial modifications to our $\chi^2$-tester.
We have already given some details on the required adaptations to the tester in Remark~\ref{rmk}.

Our algorithms for learning these classes use convex programming.
One of the main challenges is to enforce log-concavity of the PDF when learning $\cLCD$ (respectively, of the CDF when learning $\cMHR$), while simultaneously enforcing closeness in total variation distance.
This involves a careful choice of our variables, and we exploit
structural properties of the classes to ensure the soundness of
particular Taylor approximations. We encourage the reader to refer to the proofs of
Theorems~\ref{thm:unimodality},~\ref{thm:lcd-main},and~\ref{thm:mhr-main}
for more details.

\section{Testing Unimodality}
\label{sec:unimodal-main}
One striking feature of Birg\'e's result is that the decomposition of the domain is oblivious to the samples, and therefore to the unknown distribution. However, such an oblivious decomposition will not work for the unimodal distribution, since the mode is unknown. Suppose we know where the mode of the unknown distribution might be, then the problem can be decomposed into monotone functions over two intervals. Therefore, in theory, one can modify the monotonicity testing algorithm by iterating over all the possible $\absz$ modes. Indeed, by applying a union bound, it then follows that 
\begin{theorem}(Follows from Monotone)
For $\eps>1/n^{1/4}$, there exists an algorithm for testing unimodality over $[n]$ with sample complexity $O\left(\frac{\sqrt{n}}{\eps^2}\log \absz\right)$.
\end{theorem}

However, this is unsatisfactory, since our lower bound (and as we will demonstrate, the true complexity of this problem) is $\sqrt{n}/\ve^2$.
We overcome the logarithmic barrier introduced by the union bound, by
employing a non-oblivious decomposition of the domain, and using 
Kolmogorov's max-inequality. 

Our main result for testing unimodality is the following theorem,
which is proved in Section~\ref{sec:unimodal-appendix}.
\begin{theorem}
\label{thm:unimodality}
Suppose $\eps>n^{-1/4}$.
Then there exists an algorithm for testing unimodality over $[n]$ with
sample complexity $O(\sqrt{\absz}/\eps^2)$. 
\end{theorem}

%% file: independence.tex
\section{Testing Independence of Random Variables}
\label{sec:independence}
Let 
$\cX\ed[n_1]\times\ldots\times[n_d]$, and let $\prdd$ be the class
of all product distributions over $\cX$. 
We first bound the $\chi^2$-distance between product distributions in
terms of the individual coordinates. 
\begin{lemma}
\label{lem:prod-chi}
Let $\dP = \dP^1\times \dP^2\ldots\times \dP^d$, and 
$\dQ = \dQ^1\times \dQ^2\ldots\times \dQ^d$ be two distributions in $\prdd$. 
Then 
\[
\chisqpq{\dP}{\dQ} = \prod_{\gnote{\ell}=1}^d(1+\chisqpq{\dP\gnote{^\ell}}{\dQ\gnote{^\ell}})-1.
\]
\end{lemma}
\begin{proof}
By the definition of $\chi^2$-distance
\begin{align}
\chisqpq{\dP}{\dQ} &= \sum_{\bi\in\cX} \frac{\dPbi^2}{\dQbi}-1\\
& =
\prod_{\gnote{\ell}=1}^d\left[\sum_{i \in[n_\gnote{\ell}]}\frac{\left(\dP^{\gnote{\ell}}_{i}\right)^2}{\dQ^{\gnote{\ell}}_{i}}\right]-1\\
& =
\prod_{\gnote{\ell = 1}}^\gnote{d}\left(1+\chisqpq{\dP^\gnote{\ell}}{\dQ^\gnote{\ell}}\right)-1.
\end{align}
\end{proof}

Along the lines of learning monotone distributions in $\chi^2$
distance we obtain the following result, proved in Section~\ref{sec:independence-appendix}.  
\begin{lemma}
\label{lem:learning-product}
There is an algorithm that takes 
\[
O\left(\sum_{\ell=1}^d \frac{n_\ell}{\eps^2}\right)
\]
samples from a distribution $\dP$ in $\prdd$ and outputs a distribution
$\dQ\in\prdd$ such that with probability at least $5/6$, 
\[
\chisqpq{\dP}{\dQ}\le O(\eps^2).
\] 
\end{lemma}
This fits precisely in our framework of robust $\chi^2$-$\ell_1$
testing. In particular, applying Theorem~\ref{thm:chisq-test}, we obtain the following result. 
\begin{theorem}
\label{thm:independence}
For any $d\ge1$, there exists an algorithm for testing independence of
random variables over $[n_1]\times\ldots[n_d]$ with sample and time complexity 
\[
O\left(\frac{(\prod_{\ell = 1}^dn_\ell)^{1/2} + \sum_{\ell = 1}^d n_\ell}{\eps^2}\right).
\] 
\end{theorem}

The following corollaries are immediate. 
\begin{corollary}
Suppose $\prod_{\ell = 1}^d n_\ell^{1/2} \geq \sum_{\ell =1}^d n_\ell$.
Then there exists an algorithm for testing independence over $[n_1] \times \dots \times [n_d]$ with sample complexity $\Theta((\prod_{\ell = 1}^d n_\ell)^{1/2}/\eps^2)$. 
\end{corollary}

In particular, 
\begin{corollary}
There exists an algorithm for testing if two distributions over $[n]$ are independent with sample complexity $\Theta(n/\eps^2)$. 
\end{corollary}

%% file: otherclasses.tex
\section{Testing Log-Concavity}
\label{sec:LCD-main}
In this section we describe our results for testing log-concavity of
distributions. Our main result is as follows:
\begin{theorem}
\label{thm:lcd-main}
There exists an algorithm for testing log-concavity over $[n]$ with sample complexity
$$O\left(\frac{\sqrt{n}}{\eps^2}+\frac{1}{\ve^5}\right)$$
and time complexity $\poly(n,1/\ve)$.
\end{theorem}

In particular, this implies the following optimal tester for this class:
\begin{corollary}
Suppose $\ve > 1/n^{1/5}$.
Then there exists an algorithm for testing log-concavity over $[n]$ with sample complexity $O\left(\sqrt{n}/\ve^2\right)$.
\end{corollary}

Our algorithm will fit into the structure of our general framework.
We first perform a very particular type of learning algorithm, whose
guarantees are summarized in the following lemma: 
\begin{lemma}
\label{lem:lcd-learn}
Given $\ve > 0$ and sample access to a distribution $p$, there exists an algorithm with the following guarantees:
\begin{itemize}
\item If $p \in \cLCD$, the algorithm outputs a distribution $q \in \cLCD$ and an $O(\ve)$-effective support $S$ of $p$ such that $\chi^2(p_S,q_S) \leq \frac{\ve^2}{500}$ with probability at least $5/6$;
\item If $\dtv(p,\cLCD) \geq \ve$, the algorithm either outputs a
  distribution $q \in \cLCD$ or \reject. 
\end{itemize}
The sample complexity is $O(1/\ve^5)$ and the time complexity is $\poly(n,1/\ve)$.
\end{lemma}

We note that as a corollary, one immediately obtains a $O(1/\ve^5)$ proper learning algorithm for log-concave distributions.
The result is immediate from the first item of Lemma~\ref{lem:lcd-learn} and Proposition~\ref{prop:distance-relations}. 
We can actually do a bit better -- in the proof of Lemma~\ref{lem:lcd-learn}, we partition $[n]$ into intervals of probability mass $\Theta(\ve^{3/2})$.
If one instead partitions into intervals of probability mass $\Theta(\ve/\log(1/\ve))$ and works directly with total variation distance instead of $\chi^2$ distance, one can show that $\tilde O(1/\ve^4)$ samples suffice.
\begin{corollary}
\label{cor:learning LCD}
Given $\ve > 0$ and sample access to a distribution $p \in \cLCD$, there exists an algorithm which outputs a distribution $q \in \cLCD$ such that $\dtv(p,q) \leq \ve$.
The sample complexity is $\tilde O(1/\ve^4)$ and the time complexity is $\poly(n,1/\ve)$.
\end{corollary}

Then, given the guarantees of Lemma \ref{lem:lcd-learn},
Theorem~\ref{thm:lcd-main} follows from
Theorem~\ref{thm:chisq-test}\footnote{To be more precise, we require
  the modification of Theorem~\ref{thm:lcd-main} which is described in
  Section~\ref{sec:testing}, in order to handle the case where the
  $\chi^2$-distance guarantees only hold for a known effective
  support.}. The details of these results are presented in Section~\ref{sec:LCD}.

\section{Testing for Monotone Hazard Rate}
\label{sec:MHR-main}
In this section, we obtain our main result for testing for monotone hazard rate:
\begin{theorem}
\label{thm:mhr-main}
There exists an algorithm for testing monotone hazard rate over $[n]$ with sample complexity
$$O\left(\frac{\sqrt{n}}{\eps^2}+\frac{\log(n/\ve)}{\ve^4}\right)$$
and time complexity $\poly(n,1/\ve)$.
\end{theorem}

This implies the following optimal tester for the class:
\begin{corollary}
Suppose $\ve > \sqrt{\log(n/\ve)}/n^{1/4}$.
Then there exists an algorithm for testing monotone hazard rate over $[n]$ with sample complexity $O\left(\sqrt{n}/\ve^2\right)$.
\end{corollary}

We obey the same framework as before, first applying a $\chi^2$-learner with the following guarantees:
\begin{lemma}
\label{lem:mhr-learn}
Given $\ve > 0$ and sample access to a distribution $p$, there exists an algorithm with the following guarantees:
\begin{itemize}
\item If $p \in \cMHR$, the algorithm outputs a distribution $q \in \cMHR$ and an $O(\ve)$-effective support $S$ of $p$ such that $\chi^2(p_S,q_S) \leq \frac{\ve^2}{500}$ with probability at least $5/6$;
\item If $\dtv(p,\cMHR) \geq \ve$, the algorithm either outputs a distribution $q \in \cMHR$ and a set $S \subseteq [n]$ or \reject.
\end{itemize}
The sample complexity is $O(\log(n/\ve)/\ve^4)$ and the time complexity is $\poly(n,1/\ve)$.
\end{lemma}

As with log-concave distributions, this implies the following proper learning result:
\begin{corollary}
\label{cor:learning MHR}
Given $\ve > 0$ and sample access to a distribution $p \in \cMHR$, there exists an algorithm which outputs a distribution $q \in \cMHR$ such that $\dtv(p,q) \leq \ve$.
The sample complexity is $O(\log(n/\ve)/\ve^4)$ and the time complexity is $\poly(n,1/\ve)$.
\end{corollary}

Again, combining the learning guarantees of Lemma \ref{lem:mhr-learn}
with the appropriate variant of Theorem~\ref{thm:chisq-test}, we
obtain Theorem~\ref{thm:mhr-main}. The details of the argument and
proofs are presented in Section~\ref{sec:MHR}.

%% file: lower-bounds.tex
\section{Lower Bounds}
\label{sec:lower-bounds}
We now prove sharp lower bounds for the classes of distributions we
consider. We show that the example studied by
Paninski~\cite{Paninski08} to prove lower bounds on testing uniformity
can be used to prove lower bounds for the classes we consider.  
They consider a class $\cQ$ consisting of $2^{\absz/2}$ distributions
defined as follows. Without loss of generality assume that $\absz$ is even. For each of
the $2^{\absz/2}$ vectors $z_0z_1\ldots
z_{\absz/2-1}\in\{-1,1\}^{\absz/2}$, define a distribution $\dQ\in\cQ$ over
$[\absz]$ as follows.  
\begin{align}
\label{eqn:fcl}
\dQ_{i} =\begin{cases}
\frac{(1+z_\ell c\eps)}{\absz} & \text{ for } i = 2\ell+1\\
 \frac{(1-z_{\ell}c\eps)}{\absz}& \text{ for } i=2\ell.\\
\end{cases}
\end{align}

Each distribution in $\cQ$ has a total variation distance $c\ve/2$
from $U_n$, the uniform distribution over
$[n]$. By choosing $c$ to be an appropriate constant, Paninski~\cite{Paninski08} showed that a distribution picked
uniformly at random from $\cQ$ cannot be distinguished from $U_n$ with
fewer than $\sqrt{\absz}/\eps^2$ samples with probability at least $2/3$. 

Suppose $\cC$ is a class of distributions such that 
\begin{itemize}
\item
The uniform distribution $U_n$ is in $\cC$,
\item
For appropriately chosen $c$, $\dtv(\cC, \cQ)\ge\eps$,
\end{itemize}
then testing $\cC$ is \costasnote{not} easier than distinguishing $U_n$ from $\cQ$. 
Invoking~\cite{Paninski08} immediately implies that testing the class
$\cC$ requires $\Omega(\sqrt{\absz}/\eps^2)$ samples. 

The lower bounds for all the one dimensional distributions will follow
directly from this construction, and for testing monotonicity in
higher dimensions, we extend this construction to $d\ge1$,
appropriately. These arguments are proved in
Section~\ref{sec:lb-appendix}, leading to the following lower bounds for testing these classes: 
\begin{theorem}$ $
\label{thm:lbs}
\begin{itemize}
\item
For any $d\ge1$, any algorithm for testing monotonicity over
$[\absz]^d$ requires $\Omega(n^{d/2}/\eps^2)$ samples.  
\item
For $d\ge1$, any algorithm for testing independence over
$[n_1]\times\cdots\times[n_d]$ requires
$\Omega\left(\frac{(n_1\cdot n_2\ldots\cdot
    n_d)^{1/2}}{\eps^2}\right)$ samples. 
\item
Any algorithm for testing unimodality, log-concavity, or monotone
hazard rate over $[n]$ requires $\Omega(\sqrt{n}/\ve^2)$ samples.
\end{itemize}
\end{theorem}

%% file: testing-appendix.tex
\section{Moments of the Chi-Squared Statistic}
\label{sec:chisq-moments}
We analyze the mean and variance of the statistic
$$ Z = \sum_{i \in \mathcal{A}} \frac{(X_i - mq_i)^2 - X_i}{mq_i},$$
where each $X_i$ is independently distributed according to $\poi(\text{$m p_i$})$.

We start with the mean:
\begin{align*}
\expectation{Z} &= \sum_{i \in \mathcal{A}} \expectation{\frac{(X_i - mq_i)^2 - X_i}{mq_i}} \nonumber \\
     &= \sum_{i \in \mathcal{A}} \frac{\expectation{X_i^2} -
       2mq_i\expectation{X_i} + m^2q_i^2 - \expectation{X_i}}{mq_i} \nonumber \\ 
     &= \sum_{i \in \mathcal{A}} \frac{m^2p_i^2 + m p_i - 2m^2q_ip_i + m^2q_i^2 - m p_i}{mq_i} \nonumber \\
     &= m \sum_{i \in \mathcal{A}} \frac{(p_i - q_i)^2}{q_i} \nonumber\\
     &= m \cdot \chi^2(p_\cA,q_\cA) \nonumber
\end{align*}

Next, we analyze the variance.
Let $\lambda_i = \E{X_i}=m p_i$ and $\lambda_i' = m q_i$.
\begin{align}
\Var{Z} &= \sum_{i \in \mathcal{A}}\frac{1}{\l_i'^2}\Var{(X_i - \l_i)^2 + 2(X_i - \l_i)(\l_i - \l_i') - (X_i - \l_i)} \nonumber \\
       &= \sum_{i \in \mathcal{A}}\frac{1}{\l_i'^2}\Var{(X_i - \l_i)^2 + (X_i - \l_i)(2\l_i -2\l_i' - 1) } \nonumber\\
       &= \sum_{i \in \mathcal{A}}\frac{1}{\l_i'^2}\E{(X_i - \l_i)^4 + 2(X_i - \l_i)^3(2\l_i -2\l_i' - 1) + (X_i - \l_i)^2(2\l_i -2\l_i' - 1)^2 - \l_i^2} \nonumber\\
       &= \sum_{i \in \mathcal{A}}\frac{1}{\l_i'^2}[3\l_i^2 + \l_i + 2\l_i(2\l_i - 2\l_i' - 1) + \l_i(2\l_i - 2\l_i' - 1)^2 - \l_i^2] \nonumber\\
       &= \sum_{i \in \mathcal{A}}\frac{1}{\l_i'^2}[2\l_i^2 + \l_i + 4\l_i(\l_i - \l_i') - 2\l_i + \l_i(4(\l_i - \l_i')^2 -4(\l_i - \l_i') + 1)] \nonumber\\
       &= \sum_{i \in \mathcal{A}}\frac{1}{\l_i'^2}[2\l_i^2  + 4\l_i(\l_i - \l_i')^2 ] \nonumber\\
       &= \sum_{i \in \mathcal{A}}\left[2 \frac{p_i^2}{q_i^2}+4m\cdot\frac{p_i\cdot(p_i-q_i)^2}{q_i^2}\right] 
\end{align}
The third equality is by noting the random variable has expectation $\l_i $ and the fourth equality substitutes the values of centralized moments of the Poisson distribution.

\section{Analysis of our $\chi^2$-Test Statistic}
\label{sec:chisq-analysis}
We first prove the key lemmas in the analysis of our $\chi^2$-test.

\begin{prevproof}{Lemma}{lem:means}
The former case is straightforward from (\ref{eqn:mean}) and \ref{prp:in-chisq} of $q$.

We turn to the latter case. 
Recall that $\cA= \{i:q_i \geq \ve/50n\}$, and thus $q(\bar \cA) \leq \ve/50$.
We first show that $\dtv(p_{\cA}, q_{\cA})\geq \frac{6\eps}{25}$, where $p_{\cA}, q_{\cA}$ are
defined as above and in our slight abuse of notation we use $\dtv(p_{\cA}, q_{\cA})$ for non-probability vectors to denote $\frac12\|p_{\cA} - q_{\cA}\|_1.$

Partitioning the support into $\cA$ and $\compl{\cA}$, we have
\begin{align}
\dtv(p,q)=\dtv(p_{\cA}, q_{\cA})+\dtv(p_{\compl{\cA}}, q_{\compl{\cA}}).\label{eqn:tv-decomp}
\end{align}

We consider the following cases separately:
\begin{itemize}
\item 
{\bf $p(\compl{\cA})\le \eps/2$:} In this case, 
\begin{align}
  \dtv(p_{\compl{\cA}}, q_{\compl{\cA}}) = \frac12 \sum_{i \in \compl{\cA}} |p_i - q_i| \leq 
  \frac12 (p(\compl{\cA})+q(\compl{\cA})) \le \frac{1}{2}\left(\frac{\ve}{2} + \frac{\ve}{50}\right) = \frac{13\ve}{50}.\nonumber
\end{align}
Plugging this in~\eqref{eqn:tv-decomp}, and using the fact that $\dtv(p,q) \geq \ve$ 
shows that $\dtv(p_{\cA}, q_{\cA}) \geq \frac{6\ve}{25}$. 
\item 
{\bf $p(\compl{\cA})> \eps/2$:} In this case, by the reverse triangle inequality,  
\begin{align}
  \dtv(p_{\cA}, q_{\cA}) \geq \frac12 (q(\cA)-p(\cA)) \geq \frac12 ((1 - \ve/50) - (1 - \ve/2)) = \frac{6\ve}{25} .\nonumber
\end{align}
\end{itemize}

By the Cauchy-Schwarz inequality,
\begin{align}
  \chi^2(p_{\cA}, q_{\cA}) &\ge 4\frac{\dtv(p_{\cA},q_{\cA})^2}{q(\cA)}\nonumber\\
                           &\geq \frac{\ve^2}{5}. \nonumber
\end{align}
We conclude by recalling~\eqref{eqn:mean}.

\end{prevproof}

\begin{prevproof}{Lemma}{lem:vars}
We bound the terms of (\ref{eqn:variance}) separately, starting with the first.

\begin{align}
2\sum_{i \in \mathcal{A}} \frac{p_i^2}{q_i^2} &= 2\sum_{i \in \mathcal{A}} \left(\frac{(p_i - q_i)^2}{q_i^2} + \frac{2p_iq_i - q_i^2}{q_i^2}\right) \nonumber \\
                                             &= 2\sum_{i \in \mathcal{A}} \left(\frac{(p_i - q_i)^2}{q_i^2} + \frac{2q_i(p_i - q_i) + q_i^2}{q_i^2}\right) \nonumber\\
                                             &\leq 2n + 2\sum_{i \in \mathcal{A}} \left(\frac{(p_i - q_i)^2}{q_i^2} + 2\frac{(p_i - q_i)}{q_i}\right) \nonumber\\
                                             &\leq 4n + 4\sum_{i \in \mathcal{A}} \frac{(p_i - q_i)^2}{q_i^2} \nonumber\\
                                             &\leq 4n + \frac{200n}{\ve} \sum_{i \in \mathcal{A}} \frac{(p_i - q_i)^2}{q_i}\nonumber\\
                                             &= 4n + \frac{200n}{\ve}\frac{E[Z]}{m} \nonumber\\
                                             &\leq 4n + \frac{1}{100}\sqrt{n} E[Z]\label{eq:first-var-term-in}
\end{align}
The second inequality is the AM-GM inequality, the third inequality uses that $q_i \geq \frac{\ve}{50n}$ for all $i \in \cA$, the last equality uses \eqref{eqn:mean}, and the final inequality substitutes a value $m \geq 20000\frac{\sqrt{n}}{\ve^2}$.

The second term can be similarly bounded:
\begin{align*}
4m \sum_{i \in \mathcal{A}} \frac{p_i(p_i - q_i)^2}{q_i^2} &\leq 4m \left(\sum_{i \in \mathcal{A}} \frac{p_i^2}{q_i^2}\right)^{1/2}\left(\sum_{i \in \mathcal{A}} \frac{(p_i - q_i)^4}{q_i^2}\right)^{1/2} \\
                                                          &\leq 4m \left(4n + \frac{1}{100}\sqrt{n} E[Z] \right)^{1/2}\left(\sum_{i \in \mathcal{A}} \frac{(p_i - q_i)^4}{q_i^2}\right)^{1/2} \\
                                                          &\leq 4m \left(2\sqrt{n} + \frac{1}{10}n^{1/4} E[Z]^{1/2}\right)\left(\sum_{i \in \mathcal{A}} \frac{(p_i - q_i)^2}{q_i}\right) \\
                                                          &= \left(8\sqrt{n} + \frac{2}{5}n^{1/4} E[Z]^{1/2}\right)E[Z] \\
\end{align*}
The first inequality is Cauchy-Schwarz, the second inequality uses (\ref{eq:first-var-term-in}), the third inequality uses the monotonicity of the $\ell_p$ norms, and the equality uses~\eqref{eqn:mean}.

Combining the two terms, we get
$$\Var{Z} \leq 4n + 9\sqrt{n} \E{Z}  + \frac{2}{5}n^{1/4} \E{Z}^{3/2}  .$$

We now consider the two cases in the statement of our lemma.
\begin{itemize}
\item
When $p \in \mathcal{C}$, we know from Lemma~\ref{lem:means} that $\E{Z} \leq \frac{1}{500} m\ve^2$. Combined with a choice of $m \geq 20000 \frac{\sqrt{n}}{\ve^2}$ and the above expression for the variance, this gives:
$$\Var{Z} \leq \frac{4}{20000^2}m^2\ve^4 + \frac{9}{20000 \cdot 500}m^2\ve^4  + \frac{\sqrt{10}}{12500000}m^2\ve^4 \leq \frac{1}{500000}m^2\ve^4.$$

\item When $\dtv(p,\mathcal{C}) \geq \ve$, Lemma~\ref{lem:means} and  $m \geq 20000\frac{\sqrt{n}}{\ve^2}$ give:
$$\E{Z} \geq \frac{1}{5}m\ve^2 \geq 4000\sqrt{n}.$$

Combining this with our expression for variance we get:
$$\Var{Z} \leq \frac{4}{4000^2}\E{Z}^2 + \frac{9}{4000}\E{Z}^2 + \frac{2}{5\sqrt{4000}}\E{Z}^2 \leq \frac{1}{100}\E{Z}^2.$$
\end{itemize}
\end{prevproof}

%% file: monotone-appendix.tex
\section{Details on Testing Monotonicity}
\label{sec:monotone-appendix}
In this section, we prove the lemmas necessary for our monotonicity testing result.

\subsection{A Structural Result for Monotone Distributions on the Hypergrid}
\label{sec:hypergrid}
Birg\'e~\cite{Birge87} showed that any monotone
distribution is estimated to a total variation $\eps$ with a
$O(\log(\absz)/\eps)$-piecewise constant distribution. Moreover, the
intervals over which the output is constant is 
independent of the distribution $\dP$. This result, was strengthened
to the Kullback-Leibler divergence by~\cite{AcharyaJOS14a} to study the compression of
monotone distributions. They upper bound the KL divergence by
$\chi^2$ distance and then bound the $\chi^2$ distance. 
We extend this result to $[n]^d$.
We divide $[\absz]^d$ into $b^d$ rectangles
as follows.
Let $\{I_1, \dots, I_b\}$ be a partition of
$[\absz]$ into consecutive intervals defined as:
\begin{align*}
|I_j|
=
\begin{cases}
1  & \text{ for } 1\le j\le\frac b2, \\
\lfloor 2(1+\gamma)^{j-b/2} \rfloor & \text{ for }\frac b2< j\le b.
 \end{cases}
\end{align*}
For $\bj=(j_1,\ldots, j_d)\in[b]^d$, let $I_\bj\ed I_{j_1}\times
I_{j_2}\times\ldots\times I_{j_d}$. 

 The $\chi^2$ distance between $\dP$ and $\dPbar$
can be bounded as
\begin{align*}
\chi^2(\dP,\dPbar) 
=& \left[\sum_{\bj\in[b]^d}\sum_{\bi\in
    I_{\bj}}\frac{\dPbi^2}{\dPbarbi}\right]-1\\
\le& \left[\sum_{\bj\in[b]^d}\dPbj^+|I_\bj|\right]-1\\
\end{align*}
For $\bj=(j_1,\ldots, j_d)\in\largeset$, let $\bj^*=(j_1^*,\ldots,
j_b^*)$ be 
\begin{align*}
j_i^* =\begin{cases}
j_i& \text{ if } j_i\le b/2+1\\
j_i-1& \text{ otherwise}.
\end{cases}
\end{align*}

We bound the expression above as follows. 

Let $T \subseteq [d]$ be any subset of $d$. Suppose the size of $T$ is
$\ell$.  Let $\bar T$ be the set of all $\bj$ that satisfy $\bj_i=b/2+1$ for $i\in
T$. In other words, over the dimensions determined by $T$, the value
of the index is equal to $d/2+1$. The map $\bj\to\bj^*$ restricted to
$T$ is one-to-one, and since at most $d-\ell$ of the coordinates drop, 
\begin{align*}
|I_{\bj}| \le |I_{\bj^*}|\cdot(1+\gamma)^{d-\ell}. 
\end{align*}
Since there are $\ell$ coordinates that do not change, 
and each of them have $2(1+\gamma)$ coordinates, we obtain
\begin{align*}
\sum_{\bj\in\bar T}\dP_\bj \le& 
\ \sum_{\bj\in\bar T}\dP_{\bj^*}^-\cdot|I_{\bj}|\cdot
(2(1+\gamma))^{\ell}\cdot(1+\gamma)^{d-\ell}\\
=&\ \sum_{\bj\in\bar T}\dP_{\bj^*}^-\cdot|I_{\bj^*}| \cdot 2^{\ell} (1+\gamma)^{d}.
\end{align*}

Since the mapping is one-to-one, the probability of observing as element in
$\bar T$ is the probability of observing $b/2+1$ in $\ell$
coordinates, which is at most $(2/(b+2))^\ell$ under any monotone
distribution. Therefore,
\begin{align*}
\sum_{\bj\in\bar T}\dP_\bj \le&\left(\frac2{b+2}\right)^\ell \cdot 2^{\ell} (1+\gamma)^{d}.
\end{align*}
For any $\ell$ there are ${d\choose \ell}$ choices for $T$.
Therefore, 
\begin{align}
\chi^2(\dP,\dPbar)\le\ & \sum_{\ell=0}^d {d\choose \ell}
\left(\frac4{b+2}\right)^\ell (1+\gamma)^{d}-1\nonumber\\
=&\ (1+\gamma)^d\left(1+\frac4{b+2}\right)^d-1\nonumber\\
=&  \ \left(1+\gamma+\frac4{b+2}+\frac{4\gamma}{b+2}\right)^d-1\nonumber
\end{align}

Recall that $\gamma=2\log (\absz)/b>1/b$, implies that the expression above
is at most $(1+2\gamma)^d-1$. 
This implies Lemma~\ref{lem:mon-str}.

\subsection{Monotone Learning}

Our algorithm requires a distribution $\dQ$ satisfying the properties discussed earlier. We learn a monotone distribution from samples as follows. 

Before proving this result, we prove a general result for $\chi^2$ learning of arbitrary discrete distributions, adapting the result from~\cite{KamathOPS15}.
For a distribution $\dP$, and a partition of the domain into $b$ intervals $I_1, \ldots, I_b$, let $\dPbari= \dP(I_i)/|I_i|$ be the flattening of $\dP$ over these intervals. We saw that for monotone distributions there exists a partition of the domain such that $\dPbar$ is \emph{close} to the underlying distribution in $\chi^2$ distance. 

Suppose we are given $m$ samples from a distribution $\dP$ and a partition $I_1, \ldots, I_b$. Let $m_j$ be the number of samples that fall in $I_j$. For $i\in I_j$, let
\[
\dQi \ed \frac{1}{|I_j|}\frac{m_j+1}{m+b}.
\]

Let $S_j=\sum_{i\in I_j}\dPi^2$. The expected $\chi^2$ distance between $\dP$ and $\dQ$ can be bounded as follows. 
\begin{align}
\expectation{\chi^2(\dP,\dQ) } =& \left[\sum_{j=1}^b \sum_{i\in I_j} \sum_{\ell=0}^{m}{m\choose \ell}(\dP(I_j))^{\ell}(1-\dP(I_j))^{m-\ell}\frac{\dPi^2}{(\ell+1)/(|I_j|(m+b))}\right]-1\nonumber\\
= &\left[\frac{m+b}{m+1}\sum_{j=1}^b \frac{S_j}{\dPbar(I_j)/|I_j|} \left(\sum_{\ell=0}^{m}{m+1\choose \ell+1}(\dP(I_j))^{\ell+1}(1-\dP(I_j))^{m+1-\ell+1}\right)\right]-1 \nonumber\\
= & \left[\frac{m+b}{m+1}\sum_{j=1}^b \frac{S_j}{\dPbar(I_j)/|I_j|} \left(1-(1-\dP(I_j)^{m+1}\right)\right]-1\nonumber\\
\le & \left[\frac{m+b}{m+1}\sum_{j=1}^b \frac{S_j}{\dPbar(I_j)/|I_j|}\right]-1\nonumber\\
= & \left[\frac{m+b}{m+1}\left(\chi^2(\dP,\dPbar)+1\right)\right]-1\nonumber\\
= & \frac{m+b}{m+1}\cdot\chi^2(\dP,\dPbar)+\frac{b}{m+1}\label{eqn:chi-learn}.
\end{align}

Suppose $\gamma = O(\log (\absz)/b)$, and $b=O(d\cdot\log (\absz)/\eps^2)$. Then, by Lemma~\ref{lem:mon-str}, 
\begin{align}
\chi^2(\dP,\dPbar)\le \eps^2.
\end{align}

Combining this with~\eqref{eqn:chi-learn} gives Lemma~\ref{lem:fin-learn-mon}.

%% file: unimodal-appendix.tex
\section{Details on testing Unimodality}
\label{sec:unimodal}
\label{sec:unimodal-appendix}

Recall that to circumvent Birg\'e's decomposition, we want to decompose the interval into disjoint intervals such that the probability of each interval is about $O(1/b)$, where $b$ is a parameter, specified later. In particular we consider a decomposition of $[n]$ with the following properties:
\begin{enumerate}
\item
For each element $i$ with probability at least $1/b$, there is an $I_\ell=\{i\}$.
\item
There are at most two intervals with $\dP(I)\le 1/{2b}$.
\item
Every other interval $I$ satisfies $\dP(I)\in\left[\frac1{2b},\frac2b\right]$.
\end{enumerate}

Let $I_1, \ldots, I_L$ denote the partition of $[\absz]$ corresponding to these intervals. Note that $L= O(b)$. 
\begin{claim}
There is an algorithm that takes $O(b\log b)$ samples and outputs $I_1, \ldots, I_L$ satisfying the properties above.
\end{claim}

The first step in our algorithm is to estimate the \emph{total probability} within each of these intervals. 
In particular, 
\begin{lemma}
There is an algorithm that takes $m'=O(b\log b/\eps^2)$ samples from a distribution $\dP$, and with probability at least 9/10 outputs a distribution $\dQbar$ that is constant on each $I_L$. Moreover, for any $j$ such that $\dP(I_j)>1/2b$, 
$\dQbar(I_j)\in(1\pm\eps)\dP(I_j)$. 
\end{lemma}
\begin{proof}
Consider any interval $I_j$ with $\dP(I_j)\ge 1/2b$. The number of
samples $N_{I_j}$ that fall in that interval is distributed
$Binomial(m', \dP(I_j)$. Then by Chernoff bounds for $m'>12 b\log
b/\eps^2$,  
\begin{align}
\probof{|N_{I_j}-m'\dP(I_j)|>\eps m'\dP(I_j) }\le& 2\exp\left(\eps^2m'\dP(I_j)/2\right)\\
\le & \frac1{b^2},
\end{align}
where the last inequality uses the fact that $\dP(I_j)\ge 1/2b$. 
\end{proof}

The next step is estimate the distance of $\dQ$ from $\cU$. This is possible by a simple dynamic program, similar to the one used for monotonicity. 
If the estimated distance is more than $\eps/2$, we output \reject. 

Our next step is to remove certain intervals. This will be to ensure that when the underlying distribution is unimodal, we are able to estimate the distribution \emph{multiplicatively} over the remaining intervals. 
In particular, we do the following preprocessing step:
\begin{itemize}
\item 
$A= \emptyset$. 
\item 
For interval $I_j$,  
\begin{itemize}
\item 
If \begin{align}
\dQ(I_j) &\notin\left((1-\eps)\cdot\dQ(I_{j+1}),
  (1+\eps)\cdot\dQ(I_{j+1})\right)\ \ \text{ OR }\\
\dQ(I_j) &\notin\left((1-\eps)\cdot\dQ(I_{j-1}), (1+\eps)\cdot\dQ(I_{j-1})\right),
\end{align}
 add $I_j$ to $A$. 
\end{itemize}
\item
Add the (at most 2) intervals with mass at most $1/2b$ to $A$.
\item
Add all intervals $j$ with $\dQ(I_j)/|I_j|<\eps/50n$ to $A$
\end{itemize}

If the distribution is unimodal, we can prove the following about the set of intervals ${A^c}$. 
\begin{lemma}
If $\dP$ is unimodal then, 
\begin{itemize}
\item
$\dP(I_{A^c})\ge 1-\eps/25-1/b - O\left(\log n/(\eps b)\right).$
\item
Except \emph{at most one} interval in ${A^c}$ every other interval $I_j$ satisfies, 
\[
\frac{\dPj^+}{\dPj^-}\le (1+\eps). 
\]
\end{itemize}
\end{lemma}

If this holds, then the $\chi^2$ distance between $\dP$ and $\dQ$ constrained to $A^c$, is at most $\eps^2$. 
This lemma follows from the following result. 
\begin{lemma}
Let $C>2$. For a unimodal distribution over $[\absz]$, there are at
most $\frac{4\log(50\absz/\eps)}{C\eps}$ intervals $I_j$  that satisfy $\frac{\dPj^+}{\dPj^-}<(1+\eps/C)$. 
\end{lemma}
\begin{proof}
To the contrary, if there are more than $\frac{4\log(50\absz/\eps)}{C\eps}$ intervals, then at least half of them are on one side of the mode, however this implies that the ratio of the largest probability and smallest probability is at least $(1+\eps/C)^j$, and if $j>\frac{2\log(50\absz/\eps)}{C\eps}$, is at least $50\absz/\eps$, contradicting that we have removed all such elements. 
\end{proof}

We have one additional pre-processing step here. We compute $\dQ(A^c)$ and if it is smaller than $1-\eps/25$, we output \reject. 

Suppose there are $L'$ intervals in $A^c$. Then, except at most one interval in $L'$ we know that the $\chi^2$ distance between $\dP$ and $\dQ$ is at most $\eps^2$ when $\dP$ is unimodal, and the TV distance between $\dP$ and $\dQ$ is at least $\eps/2$ over $A^c$. We propose the following simple modification to take into account, the one interval that might introduce a high $\chi^2$ distance in spite of having a small total variation. If we knew the interval, we can simply remove it and proceed. Since we do not know where the interval lies, we do the following. 

\begin{enumerate}
\item
Let $Z_j$ be the $\chi^2$ statistic over the $i$th interval in $A^c$, computed
with $O(\sqrt{\absz}/\eps^2)$ samples. 
\item 
Let $Z_l$ be the largest among all $Z_j$'s.
\item 
If $\sum_{j, j\ne l}Z_j>\nsmp\eps^2/10$, output \reject.
\item
Output \accept. 
\end{enumerate}

The objective of removing the largest $\chi^2$ statistic is our substitute for not knowing the largest interval. 
We now prove the correctness of this algorithm. 

\paragraph{Case 1 $\dP\in UM_\absz$:} We only concentrate on the final step. The $\chi^2$ statistic over all but one interval are at most $c\cdot m\eps^2$, and the variance is bounded as before. Since we remove the largest statistic, the expected value of the new statistic is \emph{strictly dominated} by that of these intervals. Therefore, the algorithm outputs \accept with at least the same probability as if we removed the spurious interval. 

\paragraph{Case 2 $\dP\notin UM_\absz$:} 
This is the hard case to prove for unimodal distributions. We know that the $\chi^2$ statistic is large in this case, and we therefore have to prove that it remains large even after removing the largest test statistic $Z_l$. 

We invoke Kolmogorov's Maximal Inequality to this end. 
\begin{lemma}[Kolmogorov's Maximal Inequality]
For independent zero mean random variables $X_1,\ldots, X_L$ with finite variance, let $S_\ell=X_1+\ldots X_\ell$. Then for any $\lambda>0$, 
\begin{align}
\probof{\max_{1\le\ell\le L}\left|S_\ell\right|\ge\lambda}\le \frac{1}{\lambda^2}\cdot Var\left(S_L\right). 
\end{align}
\end{lemma}

As a corollary, it follows that $\probof{\max_{\ell}|X_\ell|>2\lambda}\le \frac{1}{\lambda^2}\cdot Var\left(S_L\right)$. 

In the case we are interested in, we let $X_i=Z_\ell-\expectation{Z_\ell}$. Then, similar to the computations before, and the fact that each interval has a small mass, it follows that that the variance of the summation is at most $\expectation{Z_\ell}^2/100$. Taking $\lambda = \expectation{S_L-m\eps^2/3}^2/100$, it follows that the statistic does not fall below to $\sqrt \absz$. 
This completes the proof of Theorem~\ref{thm:unimodality}.

%% file: independence-appendix.tex
\section{Learning product distributions in $\chi^2$ distance}
\label{sec:independence-appendix}
In this section we prove Lemma~\ref{lem:learning-product}. The proof
is analogous to the proof for learning monotone distributions, and
hinges on the following result of~\cite{KamathOPS15}. 
Given $m$ samples from a distribution $q$ over $n$ elements, the add-1
estimator (Laplace estimator) $q$ satisfies:
\[
\expectation{\chi^2(\dP,\dQ) } \le \frac{n}{m+1}.
\]

Now, suppose $p$ is a product distribution over
$\cX=[n_1]\times\cdots\times[n_d]$. We simply perform the add-1
estimation over each coordinate independently, giving a distribution
$q^1\times\cdots \times q^d$. Since $p$ is a product distribution the
estimates in each coordinate is independent. Therefore, a simple
application of the previous result and independence of the coordinates implies
\begin{align}
\expectation{\chi^2(\dP,\dQ) } & = 
\prod_{l=1}^{d}\left(1+ \expectation{\chi^2(\dP^l,\dQ^l) }\right)-1\nonumber\\
& \le \prod_{l=1}^{d}\left(1+ \frac{n_l}{m+1}\right)-1\nonumber\\
&\le \exp\left(\frac{\sum_l {n_l}}{m+1}\right)-1,\label{eqn:haha}
\end{align}
where~\eqref{eqn:haha} follows from $e^x\ge1+x$. 
Using $e^x\le 1+2x$ for $0\le x\le 1$, we have  
\begin{align}
\expectation{\chi^2(\dP,\dQ) } & \leq 2\frac{\sum_l {n_l}}{m+1},
\end{align}
when $m\geq\sum_l n_l$. Therefore, following an application of Markov's inequality, when $m = \Omega((\sum_l
n_l)/\eps^2)$, Lemma~\ref{lem:learning-product} is proved.

%% file: lcd-appendix.tex
\section{Details on Testing Log-Concavity}
\label{sec:LCD}
It will suffice to prove Lemma~\ref{lem:lcd-learn}. 

\begin{prevproof}{Lemma}{lem:lcd-learn}
We first draw samples from $p$ and obtain a $O(1/\ve^{3/2})$-piecewise constant
distribution $f$ by appropriately flattening the empirical
distribution. The proof is now in two parts.
In the first part, we show that if $p \in \cLCD$ then
$f$ will be close to $p$ in $\chi^2$ distance over its effective
support.  
The second part involves proper learning of $p$. We
will use a linear program on $f$ to find a distribution $q\in\cLCD$.
This distribution is such that if  $p\in\cLCD$, then $\chi^2(p,q)$ is
small, and otherwise the algorithm will either output some $q \in \cLCD$ (with
no other relevant guarantees) or \reject. 

We first construct $f$. Let $\hat p$ be the empirical distribution
obtained by sampling $O(1/\ve^5)$ samples from $p$.
By Lemma~\ref{lem:dkw}, with probability at least $5/6$, $\dk(p,\hat p) \leq \ve^{5/2}/10$.
In particular, note that $|p_i - \hat p_i| \leq \ve^{5/2}/10$. 
Condition on this event in the remainder of the proof.

Let $a$ be the minimum $i$ such that $p_i \geq \ve^{3/2}/5$, and let $b$ be
the maximum $i$ satisfying the same condition. 
Let $M = \{a, \dots, b\}$ or $\emptyset$ if $a$ and $b$ are undefined.
By the guarantee provided by the DKW inequality, $p_i \geq \ve^{3/2}/10$ for all $i \in M$. 
Furthermore, $\hat p_i \in p_i \pm \ve^{3/2}/10 \in (1 \pm \ve) \cdot p_i$.
For each $i \in M$, let $f_i = \hat p_i$. 
We note that $|M| = O(1/\ve)$, so this contributes $O(1/\ve)$ constant pieces to $f$.

We now divide the rest of the domain into $t$ intervals, all but constantly many of measure $\Theta(\ve^{3/2})$ (under $p$).
This is done via the following iterative procedure.
As a base case, set $r_0 = 0$.
Define $I_j$ as $[l_j, r_j]$, where $l_j = r_{j-1} + 1$ and $r_j$ is the largest $j \in [n]$ such that $\hat p(I_j) \leq 9\ve^{3/2}/10$.
The exception is if $I_j$ would intersect $M$ -- in this case, we ``skip'' $M$: set $r_j = a-1$ and $l_{j+1} = b+1$.
If such a $j$ exists, denote it by $j^*$. 
We note that $p(I_j) \leq \hat p(I_j) + \ve^{5/2}/10 \leq \ve^{3/2}$.
Furthermore, for all $j$ except $j^*$ and $t$, $r_j + 1 \not \in M$,
so $p(I_j) \geq 9\ve^{3/2}/10 - \ve^{3/2}/5 - \ve^{5/2}/10 \geq 3\ve^{3/2}/5$. 
Observe that this lower bound implies that $t \leq \frac{2}{\ve^{3/2}}$ for $\ve$ sufficiently small.

\paragraph{Part 1.}
For this part of the algorithm, we only care about the guarantees when
$p \in \cLCD$, so we assume this is the case. 

For the domain $[n] \setminus M$, we let $f$ be the flattening of
$\hat p$ over the intervals $I_1, \dots I_t$. 
To analyze $f$, we need a structural property of log-concave
distributions due to Chan, Diakonikolas, Servedio, and Sun
\cite{ChanDSS13a}. This essentially states that a log-concave
distribution cannot have a sudden increase in probability. 

\begin{lemma}[Lemma 4.1 in \cite{ChanDSS13a}]
\label{lem:lcd-flat}
Let $p$ be a distribution over $[n]$ that is non-decreasing and
log-concave on $[1,x] \subseteq [n]$. 
Let $I = [x,y]$ be an interval of mass $P(I) = \tau$, and suppose that 
the interval $J = [1,x-1]$ has mass $p(J) = \s > 0$. Then 
$$p(y)/p(x) \leq 1 + \tau/\s.$$
\end{lemma}

Recall that any log-concave distribution is unimodal, and suppose the
mode of $p$ is at $i_0$.
We will first focus on the intervals $I_1, \dots, I_{t_L}$ which lie entirely to the left of $i_0$ and $M$.
We will refer to $I_j$ as $L_j$ for all $j \leq t_L$.
Note that $p$ is non-decreasing over these intervals.

The next steps to the analysis are as follows.
First we show that the flattening of $p$ over $L_j$ is a multiplicative $(1 + O(1/j))$ estimate for each $p_i \in L_j$.
Then, we show that flattening the empirical distribution $\hat p$ over $L_j$ is a multiplicative $(1 + O(1/j))$ estimate of $p(i)$ for each $i \in L_j$.
Finally, we exclude a small number of intervals (those corresponding to $O(\ve)$ mass at the left and right side of the domain, as well as $j^*$) in order to get the $\chi^2$ approximation we desire on an effective support.
\begin{itemize}
\item First, recall that $p(L_j) \leq \ve^{3/2}$ for all $j$.
Also, letting $J_j = [1, r_{j-1}]$, we have that $p(J_j) \geq (j-1)\cdot 3\ve^{3/2}/5$.
Thus by Lemma \ref{lem:lcd-flat}, $p(r_j) \leq p(l_j) (1 + 2/(j-1))$.
Since the distribution is non-decreasing in $L_j$, the flattening $\bar p$ of $p$ is such that $\bar p(i) \in p(i) (1 \pm \frac{2}{j-1})$ for all $i \in L_j$.
\item We have that $p(L_j) \geq 3\ve^{3/2}/5$, and $\hat p(L_j) \in p(L_j) \pm \ve^{5/2}/10$, so $\hat p(L_j) \in p(L_j) \cdot (1 \pm \frac{\ve}{6})$, and hence
$\hat p(i) \in \bar p(i) \cdot (1 \pm \frac{\ve}{6})$ for all $i \in L_j$. 
Combining with the previous point, we have that 
$$\hat p(i) \in p(i) \cdot \left(1 \pm \left(\frac{2\ve}{3(j-1)} + \frac{\ve}{6} + \frac{2}{j-1}\right)\right) \in 
p(i) \cdot \left(1 \pm \frac{11}{3(j-1)}\right).$$
\end{itemize}

A symmetric statement holds for the intervals that lie entirely to
the right of $i_0$ and $M$. 
We will refer to $I_j$ as $R_{t-j}$ for all $j > t_L$. 

To summarize, we have the following guarantees for the distribution $f$:
\begin{itemize}
\item For all $i \in M$, $f(i) \in p(i) \cdot (1 \pm \ve)$;
\item For all $i \in L_j$ (except $L_1$ and $L_{j^*}$), $f(i) \in p(i) \cdot \left(1 \pm \frac{22}{3j}\right)$;
\item For all $i \in R_j$ (except $R_1$), $f(i) \in p(i) \cdot \left(1 \pm \frac{22}{3j}\right)$;
\end{itemize}
Note that, in particular, we have multiplicative estimates for all intervals, except those in $L_1$, $L_{j^*}$, $R_1$ and the interval containing $i_0$.
Let $S$ be the set of all intervals except $L_{j^*}$, $L_j$ and $R_j$ for $j \leq 1/\sqrt{\ve}$, and the one containing $i_0$ 
Then, since each interval has probability mass at most $O(\ve^{3/2})$ and we are excluding $O(1/\sqrt{\ve})$ intervals,
$p(S)>1-O(\ve)$. 

We now compute the $\chi^2$-distance induced by this approximation for elements in $S$. 
For an element $i \in L_j \cap S$, we have
$$\frac{(f(i) - p(i))^2}{p(i)} \leq \frac{60p(i)}{j^2}.$$
Summing over all $i \in L_j \cap S$ gives
$$\frac{60\ve^{3/2}}{j^2}$$
since the probability mass of $L_j$ is at most $\ve^{3/2}$. 
Summing this over all $L_j$ for $j \geq 1/\sqrt{\ve}$ and $j \neq j^*$ gives
\begin{align*}
  60\ve^{3/2} \sum_{j = 1/\sqrt{\ve}}^{2/\ve^{3/2}} \frac{1}{j^2} &\leq 60\ve^{3/2} \int_{1/\sqrt{\ve}}^{\infty} \frac{1}{x^2} dx \\
                                                                  &= 60\ve^{3/2}(\sqrt{\ve}) \\
                                                                  &= O(\ve^2)
\end{align*}
as desired. 
\paragraph{Part 2.} 
To obtain a distribution $q\in\cLCD$, we write a linear program. 
We will work in the log domain, so our variables will be $Q_i$,
representing $\log q(i)$ for $i \in [n]$. 
We will use $F_i = \log f(i)$ as parameters in our LP.
There will be no objective function, we simply search for a feasible point.
Our constraints will be
$$Q_{i-1} + Q_{i+1} \leq 2 Q_{i} \ \ \forall i \in [n-1]$$
$$Q_{i} \leq 0 \ \ \forall i \in [n]$$
$$ \log (1 + \ve) \leq |Q_i - F_i| \leq \log (1 + \ve) \ \text{for}\ i \in M$$
$$ \log \left(1 - \frac{22}{3j}\right) \leq |Q_i - F_i| \leq \log \left(1 + \frac{22}{3j}\right) \ \text{for}\ i \in L_j, j \geq 1/\sqrt{\ve} \ \text{and}\ j \neq j^*$$
$$ \log \left(1 - \frac{22}{3j}\right) \leq |Q_i - F_i| \leq \log \left(1 + \frac{22}{3j}\right) \ \text{for}\ i \in R_j, j \geq 1/\sqrt{\ve}$$

If we run the linear program, then after a rescaling and summing the error over all the intervals in the LP gives us that the distance between $p$ and $q$ to be $O(\ve^2)$ $\chi^2$-distance in a set $S$ which has measure $p(S) \geq 1 - 4\ve$, as desired.

If the linear program finds a feasible point, then we obtain a $q \in \cLCD$.
Furthermore, if $p \in \cLCD$, this also tells us that (after a rescaling of $\ve$), summing the error over all intervals implies that $\chi^2(p_S,q_S) \leq \frac{\ve^2}{500}$ for a known set $S$ with $p(S) \geq 1 - O(\ve)$, as desired.
If $M \neq \emptyset$, this algorithm works as described.
The issue is if $M = \emptyset$, then we don't know when the $L$ intervals end and the $R$ intervals begin.
In this case, we run $O(1/\ve)$ LPs, using each interval as the one containing $i_0$, and thus acting as the barrier between the $L$ intervals (to its left) and the $R$ intervals (to its right).
If $p$ truly was log-concave, then one of these guesses will be correct and the corresponding LP will find a feasible point.
\end{prevproof}

%% file: mhr-appendix.tex
\section{Details on MHR testing}
\label{sec:MHR}
\begin{prevproof}{Lemma}{lem:mhr-learn}
As with log-concave distributions, our method for MHR distributions can be split into two parts.
In the first step, if $p \in \cMHR$, we obtain a distribution $q$ which is $O(\ve^2)$-close to $p$ in $\chi^2$ distance on a set $\mathcal{A}$ of intervals such that $p(\mathcal{A}) \geq 1 - O(\ve)$.
$q$ will achieve this by being a multiplicative $(1 + O(\ve))$ approximation for each element within these intervals.
This step is very similar to the decomposition used for unimodal distributions (described in Section~\ref{sec:unimodal-appendix}), so we sketch the argument and highlight the key differences.

The second step will be to find a feasible point in a linear program.
If $p \in \cMHR$, there should always be a feasible point, indicating that $q$ is close to a distribution in $\cMHR$ (leveraging the particular guarantees for our algorithm for generating $q$).
If $\dtv(p,\cMHR) \geq \ve$, there may or may not be a feasible point, but when there is, it should imply the existence of a distribution $p^* \in \cMHR$ such that $\dtv(q,p^*) \leq \ve/2$. 

The analysis will rely on the following lemma from \cite{ChanDSS13a}, which roughly states that an MHR distribution is ``almost'' non-decreasing.
\begin{lemma}[Lemma 5.1 in \cite{ChanDSS13a}]
\label{lem:mhr-str}
Let $p$ be an MHR distribution over $[n]$.
Let $I = [a,b] \subset [n]$ be an interval, and $R = [b+1,n]$ be the elements to the right of $I$.
Let $\eta = p(I)/p(R)$. Then $p(b+1) \geq \frac{1}{1 + \eta}p(a)$. 
\end{lemma}

\paragraph{Part 1.}
As before, with unimodal distributions, we start by taking $O(\frac{b \log b}{\ve^2})$ samples, with the goal of partitioning the domain into intervals of mass approximately $\Theta(1/b)$.
First, we will ignore the left and rightmost intervals of mass $\Theta(\ve)$.
For all ``heavy'' elements with mass $\geq \Theta(1/b)$, we consider them as singletons.
We note that Lemma~\ref{lem:mhr-str} implies that there will be at most $O(1/\ve)$ contiguous intervals of such elements.
The rest of the domain is greedily divided (from left to right) into intervals of mass $\Theta(1/b)$, cutting an interval short if we reach one of the heavy elements.
This will result in the guarantee that all but potentially $O(1/\ve)$ intervals have $\Theta(1/b)$ mass.

Next, similar to unimodal distributions, considering the flattened distribution, we discard all intervals for which the per-element probability is not within a $(1 \pm O(\ve))$ multiplicative factor of the same value for both neighboring intervals.
The claim is that all remaining intervals will have the property that the per-element probability is within a $(1 \pm O(\ve))$ multiplicative factor of the true probability.
This is implied by Lemma~\ref{lem:mhr-str}. 
If there were a point in an interval which was above this range, the distribution must decrease slowly, and the next interval would have a much larger per-element weight, thus leading to the removal of this interval.
A similar argument forbids us from missing an interval which contains a point that lies outside this range.
Relying on the fact that truncating the left and rightmost intervals eliminates elements with low probability mass, similar to the unimodal case, one can show that we will remove at most $\log(n/\ve)/\ve$ intervals, and thus a $\log(n/\ve)/b\ve$ probability mass. 
Choosing $b = \Omega(\ve^2/\log(n/\ve))$ limits this to be $O(\ve)$, as desired.
At this point, if $p$ is indeed MHR, the multiplicative estimates guarantee that the result is $O(\ve^2)$-close in $\chi^2$-distance among the remaining intervals.

\paragraph{Part 2.}
We note that an equivalent condition for distribution $f$ being MHR is log-concavity of $\log(1 - F)$, where $F$ is the CDF of $f$.
Therefore, our approach for this part will be similar to the approach used for log-concave distributions.

Given the output distribution $q$ from the previous part of this algorithm, our goal will be check if there exists an MHR distribution $f$ which is $O(\ve)$-close to $q$.
We will run a linear program with variables $\mathfrak{f}_i = \log(1 - F_i)$.
First, we ensure that $f$ is a distribution.
This can be done with the following constraints:
\begin{alignat*}{3}
&\mathfrak{f}_i &&\leq 0 \hphantom{space}&&\forall i \in [n] \\
&\mathfrak{f}_i &&\geq \mathfrak{f}_{i+1} &&\forall i \in [n-1] \\
&\mathfrak{f}_n &&= -\infty
\end{alignat*}
To ensure that $f$ is MHR, we use the following constraint:
\begin{alignat*}{3}
&\mathfrak{f}_{i-1} + \mathfrak{f}_{i+1}  &&\leq 2\mathfrak{f}_i \hphantom{space}&&\forall i \in [2,n-1]
\end{alignat*}
Now, ideally, we would like to ensure $f$ and $q$ are $\ve$-close in total variation distance by ensuring they are pointwise within a multiplicative $(1 \pm \ve)$ factor of each other:
$$(1 - \ve) \leq f_i/q_i \leq (1 + \ve)$$
We note that this is a stronger condition than $f$ and $q$ being $\ve$-close, but if $p \in \cMHR$, the guarantees of the previous step would imply the existence of such an $f$.

We have a separate treatment for the identified singletons (i.e., those with probability $\geq 1/b$) and the remainder of the support.
For each element $q_i$ identified to have $\geq 1/b$ mass, we add two constraints:
$$\log((1-\ve/2b)(1 - Q_i)) \leq \mathfrak{f}_i \leq \log((1+\ve/2b)(1 - Q_i))$$
$$\log((1-\ve/2b)(1 - Q_{i-1})) \leq \mathfrak{f}_{i-1} \leq \log((1+\ve/2b)(1 - Q_{i-1}))$$
If we satisfy these constraints, it implies that
$$q_i - \ve/b \leq f_{i} \leq q_i + \ve/b.$$
Since $q_i \geq 1/b$, this implies $$(1 - \ve)q_i \leq f_i \leq (1 + \ve)q_i$$
as desired. 

Now, the remaining elements each have $\leq 1/b$ mass.
For each such element $q_i$, we create a constraint
$$(1 - O(\ve))\frac{q_i}{1 - Q_{i-1}} \leq \mathfrak{f}_{i-1} - \mathfrak{f}_i  \leq (1 + O(\ve))\frac{q_i}{1 - Q_{i-1}} $$
Note that the middle term is
$$-\log\left(\frac{1 - F_i}{1 - F_{i-1}}\right) = -\log\left(1 - \frac{f_i}{1 - F_{i-1}}\right) \in \frac{f_i}{1 - F_{i-1}}\left(1 \pm 2\ve\right),$$
where the second equality uses the Taylor expansion and the facts that $f_i \leq 1/b$ and $1 - F_{i-1} \geq \ve$ (since during the previous part, we ignored the rightmost $O(\ve)$ probability mass).
If we satisfy the desired constraints, it implies that
$$f_i \in \frac{1}{\left(1 \pm 2\ve\right)}\frac{1 - F_{i-1}}{1 - Q_{i-1}}(1 \mp O(\ve))q_i.$$
Since we are taking $\Omega(1/\ve^4)$ samples and $1 - F_{i-1} \geq \Omega(\ve)$, Lemma~\ref{lem:dkw} implies that $f_i$ is indeed a multiplicative $(1 \pm \ve)$ approximation for these points as well.

We note that all points which do not fall into these two cases make up a total of $O(\ve)$ probability mass.
Therefore, $f$ may be arbitrary at these points and only incur $O(\ve)$ cost in total variation distance.

If we find a feasible point for this linear program, it implies the existence of an MHR distribution within $O(\ve)$ total variation distance.
In this case, we continue to the testing portion of the algorithm.
Furthermore, if $p \in \cMHR$, our method for generating $q$ certifies that such a distribution exists, and we continue on to the testing portion of the algorithm.
\end{prevproof}

%% file: lb-appendix.tex
\section{Details of the Lower Bounds}
\label{sec:lb-appendix}
In this section, for the class of distributions $\cQ$ described in discussion on lower bounds and a class of interest $\cC$, we show that $\dtv(\cC,\cQ) \geq \ve$, thus implying a lower bound of $\Omega(\sqrt{n}/\ve^2)$ for testing $\cC$. 
\subsection{Monotone distributions}
We first consider $d=1$ and prove that for appropriately chosen $c$, any monotone distribution over $[\absz]$ is $\eps$-far from  all distributions in $\cQ$.  
Consider any $q \in \cQ$.
For this distribution, we say that $i\in[\absz]$ is a \emph{raise-point} if $\dQi<\dQipo$. Let $R_{\dQ}$ be the set of raise points of $\dQ$. For $\dQ\in\cQ$,~\eqref{eqn:fcl} implies at least one in every four consecutive integers in $[n]$ is a raise point, and therefore, $|R_\dQ|\ge\absz/4$. Moreover, note that if $i$ is a raise-point, then $i+1$ is not a raise point. For any monotone (decreasing) distribution $\dP$, $\dPi\ge\dPipo$. For any raise-point $i\in R_\dQ$, by the triangle inequality,
\begin{align}
\label{eqn:triangle-mon}
|\dPi-\dQi|+|\dPipo-\dQipo|\ge |\dPi-\dPipo+\dQipo-\dQi|\ge \dQipo-\dQi = \frac{2c\eps}{\absz}. 
\end{align}
Summing over the set $R_\dQ$, we obtain $\dtv(\dP,\dQ)\ge \frac12|R_\dQ|\cdot \frac{2c\eps}{\absz}\ge {c\eps}/{4}$. Therefore, if $c\ge4$, then $\dtv(\mathcal{M}_{n},\dQ)\ge\eps$. This proves the lower bound for $d=1$. 

This argument can be extended to $[\absz]^d$. Consider the following class of distributions on $[\absz]^d$. For each point $\bi=(i_1, \ldots, i_d)\in[\absz]^d$, where $i_1$ is even, generate a random $z\in\{-1,1\}$, and assign 
to $\bi$ a probability of $(1+z c\eps)/\absz^{d}$. Let $\be_1\ed(1,0,\ldots,0)$.  Similar to $d=1$, assign a probability $(1-z c\eps)/\absz^{d}$ to the point $\bi+\be_1 = (i_1+1,i_2,  \ldots, i_d)$. This class consists of $2^{\frac{\absz^{d/2}}2}$ distributions, and Paninski's arguments extend to give a lower bound of $\Omega(n^{d/2}/\eps^2)$ samples to distinguish this class from the uniform distribution over $[\absz]^d$. It remains to show that all these distributions are $\eps$ far from $\cM$. Call a point $\bi$ as a raise point if $\dP_\bi<\dP_{\bi+\be_1}$. For any $\bi$, one of the points $\bi$, $\bi+\be_1$, $\bi+2\be_1$, $\bi+3\be_1$ is a raise point, and the number of raise points is at least $n^{d}/4$. Invoking the triangle inequality (identical to~\eqref{eqn:triangle-mon}) over the raise-points, in the first dimension shows that any monotone distribution over $[\absz]^d$ is at a distance $\frac{c\eps}4$ from any distribution in this class. Choosing $c=4$ yields a  bound of $\eps$. 

\subsection{Testing Product Distributions}
Our idea for testing independence is similar to the previous section. We sketch the construction of a class of distributions on $\cX=[n_1]\times\cdots\times[n_d]$. Then $|\cX| = n_1\cdot n_2\ldots\cdot n_d$. For each element in $\cX$ assign a value $(1\pm c\eps)$ and then for each such assignment, normalize the values so that they add to 1, giving rise to a distribution. This gives us a class of $2^{|\cX|}$ distributions. The key argument is to show that a \emph{large} fraction of these distributions are far from being a product distribution. This follows since the degrees of freedom of a product distribution is exponentially smaller than the number of possible distributions. The second step is to simply apply Paninski's argument, now over the larger set of distributions, where we show that distinguishing the collection of distributions we constructed from the uniform distribution over $\cX$ (which is a product distribution) requires $\sqrt{|\cX|}/\eps^2$ samples.

\subsection{Log-concave and Unimodal distributions}
We will show that any log-concave or unimodal distribution is $\ve$-far from all distributions in $\cQ$. 
Since $\cLCD \subset \cU$, it will suffice to show this for every unimodal distribution. 
Consider any unimodal distribution $\dP$, with mode $\ell$. Then, $\dP$ is monotone non-decreasing over the interval $[\ell]$ and non-increasing over $\{\ell+1, \ldots, \absz\}$.  By the argument for monotone distributions, the total variation distance between $\dP$ and any distribution $\dQ$ over elements greater than $\ell$ is at least $\frac{\absz-\ell-1}{\absz}\frac{c\eps}{4}$, and over elements less than $\ell$ is at least $\frac{\ell-1}{\absz}\frac{c\eps}{4}$. Summing these two gives the desired bound.

\subsection{Monotone Hazard distributions}
We will show that any monotone hazard rate distribution is $\ve$-far from all distributions in $\cQ$. 

Let $\dP$ be any monotone-hazard distribution. Any distribution $\dQ\in\cQ$ has mass at least $1/2$ over the interval $I=[\absz/4,3\absz/4]$. Therefore, by Lemma~\ref{lem:mhr-str}, for any $i\in I$, $\dPipo\left(1+\frac{\dPi}{1/4}\right)\ge \dPi$. 
As noted before, at least $\absz/8$ of the raise-points are in  $I$. 

For any $i\in I\cap R_\dQ$, $\dQi=(1+c\eps)/\absz$, $\dQipo=(1-c\eps)/\absz$
\begin{align}
d_i = |\dPi-\dQi|+|\dPipo-\dQipo|.
\end{align}

If $\dPi\ge(1+2c\eps)/\absz$ or $\dPi\le1/\absz$, then the first term, and therefore $d_i$ is at least $c\eps/\absz$. 
If $\dPi\in(1/\absz, (1+2c\eps)/\absz)$, then for $\absz>5/(c\eps)$
\[
\dPipo\ge \frac{1}{\absz}\cdot \frac{1}{1+\frac4{\absz}}\ge \frac{1-c\eps/2}{\absz}. 
\]
Therefore the second term of $d_i$ is at $c\eps/2\absz$. Since there are at least $\absz/8$ raise points in $I$,   
\begin{align}
\dtv(\dP,\dQ)\ge \frac12\frac{\absz}{8}\cdot \frac{c\eps}{2\absz}\ge \frac{c\eps}{16}. 
\end{align}
Thus any MHR distribution is $\ve$-far from $\cQ$ for $c\ge16$. 

%% file: main-long.bbl
\newcommand{\etalchar}[1]{$^{#1}$}
\newcommand{\noopsort}[1]{} \newcommand{\printfirst}[2]{#1}
  \newcommand{\singleletter}[1]{#1} \newcommand{\switchargs}[2]{#2#1}
\begin{thebibliography}{CDGR15b}

\bibitem[AAK{\etalchar{+}}07]{alon2007testing}
Noga Alon, Alexandr Andoni, Tali Kaufman, Kevin Matulef, Ronitt Rubinfeld, and
  Ning Xie.
\newblock Testing k-wise and almost k-wise independence.
\newblock In {\em Proceedings of STOC}, 2007.

\bibitem[ACS10]{ACS10}
Michal Adamaszek, Artur Czumaj, and Christian Sohler.
\newblock Testing monotone continuous distributions on high-dimensional real
  cubes.
\newblock In {\em SODA}, pages 56--65, 2010.

\bibitem[AD15]{AcharyaD15}
Jayadev Acharya and Constantinos Daskalakis.
\newblock Testing {P}oisson {B}inomial {D}istributions.
\newblock In {\em Proceedings of SODA}, pages 1829--1840, 2015.

\bibitem[ADJ{\etalchar{+}}12]{AcharyaDJOPS12}
Jayadev Acharya, Hirakendu Das, Ashkan Jafarpour, Alon Orlitsky, Shengjun Pan,
  and Ananda~Theertha Suresh.
\newblock Competitive classification and closeness testing.
\newblock In {\em COLT}, pages 22.1--22.18, 2012.

\bibitem[ADLS15]{AcharyaDLS15}
Jayadev Acharya, Ilias Diakonikolas, Jerry Li, and Ludwig Schmidt.
\newblock Sample-optimal density estimation in nearly-linear time.
\newblock {\em arXiv preprint arXiv:1506.00671}, 2015.

\bibitem[AJOS14]{AcharyaJOS14a}
Jayadev Acharya, Ashkan Jafarpour, Alon Orlitsky, and Ananda~Theertha Suresh.
\newblock Efficient compression of monotone and $m$-modal distributions.
\newblock In {\em Proceedings of the 2014 IEEE International Symposium on
  Information Theory}, ISIT '14, pages 1867--1871, Washington, DC, USA, 2014.
  IEEE Computer Society.

\bibitem[AJOT13]{AcharyaJOS13}
Jayadev Acharya, Ashkan Jafarpour, Alon Orlitsky, and Ananda {Theertha Suresh}.
\newblock A competitive test for uniformity of monotone distributions.
\newblock In {\em Proceedings of AISTATS}, pages 57--65, 2013.

\bibitem[AK11]{agresti2011categorical}
Alan Agresti and Maria Kateri.
\newblock {\em Categorical data analysis}.
\newblock Springer, 2011.

\bibitem[BBBB72]{BBBB:72}
R.~E. Barlow, D.~J. Bartholomew, J.~M. Bremner, and H.~D. Brunk.
\newblock {\em Statistical Inference under Order Restrictions}.
\newblock Wiley, New York, 1972.

\bibitem[BFF{\etalchar{+}}01]{batu2001testing}
Tugkan Batu, Eldar Fischer, Lance Fortnow, Ravi Kumar, Ronitt Rubinfeld, and
  Patrick White.
\newblock Testing random variables for independence and identity.
\newblock In {\em Proceedings of FOCS}, 2001.

\bibitem[BFRV11]{Bhattacharyya11}
Arnab Bhattacharyya, Eldar Fischer, Ronitt Rubinfeld, and Paul Valiant.
\newblock Testing monotonicity of distributions over general partial orders.
\newblock In {\em ICS}, pages 239--252, 2011.

\bibitem[Bir87]{Birge87}
Lucien Birg{\'e}.
\newblock Estimating a density under order restrictions: Nonasymptotic minimax
  risk.
\newblock {\em The Annals of Statistics}, 15(3):995--1012, September 1987.

\bibitem[BJR11]{BJR11}
Fadoua Balabdaoui, Hanna Jankowski, and Kaspar Rufibach.
\newblock Maximum likelihood estimation and confidence bands for a discrete
  log-concave distribution, 2011.

\bibitem[BKR04]{BatuKR04}
Tu\u{g}kan Batu, Ravi Kumar, and Ronitt Rubinfeld.
\newblock Sublinear algorithms for testing monotone and unimodal distributions.
\newblock In {\em Proceedings of the 36th Annual ACM Symposium on the Theory of
  Computing}, STOC '04, New York, NY, USA, 2004. ACM.

\bibitem[BW10]{BW10sn}
Fadoua Balabdaoui and Jon~A. Wellner.
\newblock Estimation of a $k$-monotone density: characterizations, consistency
  and minimax lower bounds.
\newblock {\em Statistica Neerlandica}, 64(1):45--70, 2010.

\bibitem[Can15]{canonne2015survey}
Cl{\'e}ment~L Canonne.
\newblock A survey on distribution testing: your data is big, but is it blue.
\newblock In {\em Electronic Colloquium on Computational Complexity (ECCC)},
  volume~22, page~7, 2015.

\bibitem[CDGR15a]{CanonneDGR15a}
Clement Canonne, Ilias Diakonikolas, Themis Gouleakis, and Ronitt Rubinfeld.
\newblock Personal communication, February 2015.

\bibitem[CDGR15b]{CanonneDGR15b}
Clement Canonne, Ilias Diakonikolas, Themis Gouleakis, and Ronitt Rubinfeld.
\newblock Testing shape restrictions of discrete distributions.
\newblock In {\em STACS}, 2015.

\bibitem[CDSS13]{ChanDSS13a}
Siu~On Chan, Ilias Diakonikolas, Rocco~A. Servedio, and Xiaorui Sun.
\newblock Learning mixtures of structured distributions over discrete domains.
\newblock In {\em Proceedings of the 24th Annual ACM-SIAM Symposium on Discrete
  Algorithms}, SODA '13, pages 1380--1394, Philadelphia, PA, USA, 2013. SIAM.

\bibitem[CDSS14]{ChanDSS13b}
Siu~On Chan, Ilias Diakonikolas, Rocco~A. Servedio, and Xiaorui Sun.
\newblock Efficient density estimation via piecewise polynomial approximation.
\newblock In {\em Proceedings of the 46th Annual ACM Symposium on the Theory of
  Computing}, STOC '14, New York, NY, USA, 2014. ACM.

\bibitem[CDVV14]{ChanDVV13}
Siu{-}On Chan, Ilias Diakonikolas, Gregory Valiant, and Paul Valiant.
\newblock Optimal algorithms for testing closeness of discrete distributions.
\newblock In {\em Proceedings of SODA}, pages 1193--1203. Society for
  Industrial and Applied Mathematics (SIAM), 2014.

\bibitem[CS10]{cule2010theoretical}
Madeleine Cule and Richard Samworth.
\newblock Theoretical properties of the log-concave maximum likelihood
  estimator of a multidimensional density.
\newblock {\em Electronic Journal of Statistics}, 4:254--270, 2010.

\bibitem[DKW56]{DvoretzkyKW56}
A.~Dvoretzky, J.~Kiefer, and J.~Wolfowitz.
\newblock Asymptotic minimax character of the sample distribution function and
  of the classical multinomial estimator.
\newblock {\em The Annals of Mathematical Statistics}, 27(3):642--669, 09 1956.

\bibitem[Fis25]{Fisher25}
Ronald~Aylmer Fisher.
\newblock {\em Statistical Methods for Research Workers}.
\newblock Oliver and Boyd, Edinburgh, 1925.

\bibitem[Fis01]{Fischer01}
Eldar Fischer.
\newblock The art of uninformed decisions: A primer to property testing.
\newblock {\em Science}, 75:97--126, 2001.

\bibitem[Gol98]{Goldreich98}
Oded Goldreich.
\newblock Combinatorial property testing (a survey).
\newblock In {\em In: Randomization Methods in Algorithm Design}, pages 45--60.
  American Mathematical Society, 1998.

\bibitem[GS02]{GibbsS02}
Alison~L. Gibbs and Francis~E. Su.
\newblock On choosing and bounding probability metrics.
\newblock {\em International Statistical Review}, 70(3):419--435, dec 2002.

\bibitem[HVK05]{hall2005testing}
Peter Hall and Ingrid Van~Keilegom.
\newblock Testing for monotone increasing hazard rate.
\newblock {\em Annals of Statistics}, pages 1109--1137, 2005.

\bibitem[JW09]{JW:09}
Hanna~K. Jankowski and Jon~A. Wellner.
\newblock Estimation of a discrete monotone density.
\newblock {\em Electronic Journal of Statistics}, 3:1567--1605, 2009.

\bibitem[KOPS15]{KamathOPS15}
Sudeep Kamath, Alon Orlitsky, Dheeraj Pichapati, and Ananda~T. Suresh.
\newblock On learning distributions from their samples.
\newblock In {\em COLT}, 2015.

\bibitem[LR06]{lehmann2006testing}
Erich~L Lehmann and Joseph~P Romano.
\newblock {\em Testing statistical hypotheses}.
\newblock Springer Science \& Business Media, 2006.

\bibitem[LRR13]{levi2013testing}
Reut Levi, Dana Ron, and Ronitt Rubinfeld.
\newblock Testing properties of collections of distributions.
\newblock {\em Theory of Computing}, 9(8):295--347, 2013.

\bibitem[Mas90]{Massart90}
Pascal Massart.
\newblock The tight constant in the {D}voretzky-{K}iefer-{W}olfowitz
  inequality.
\newblock {\em The Annals of Probability}, 18(3):1269--1283, 07 1990.

\bibitem[Pan08]{Paninski08}
Liam Paninski.
\newblock A coincidence-based test for uniformity given very sparsely sampled
  discrete data.
\newblock {\em IEEE Transactions on Information Theory}, 54(10), 2008.

\bibitem[Ron08]{Ron08}
Dana Ron.
\newblock Property testing: {A} learning theory perspective.
\newblock {\em Foundations and Trends in Machine Learning}, 1(3):307--402,
  2008.

\bibitem[RS81]{rao1981analysis}
Jon~NK Rao and Alastair~J Scott.
\newblock The analysis of categorical data from complex sample surveys:
  chi-squared tests for goodness of fit and independence in two-way tables.
\newblock {\em Journal of the American Statistical Association},
  76(374):221--230, 1981.

\bibitem[Rub06]{Rubinfeld06}
Ronitt Rubinfeld.
\newblock Sublinear-time algorithms.
\newblock In {\em International Congress of Mathematicians}, 2006.

\bibitem[SW14]{SumardW14}
Adrien Saumard and Jon~A Wellner.
\newblock Log-concavity and strong log-concavity: a review.
\newblock {\em Statistics Surveys}, 8:45--114, 2014.

\bibitem[VV11]{ValiantV11}
Gregory Valiant and Paul Valiant.
\newblock Estimating the unseen: An $n/\log n$-sample estimator for entropy and
  support size, shown optimal via new {CLT}s.
\newblock In {\em Proceedings of the 43rd Annual ACM Symposium on the Theory of
  Computing}, STOC '11, pages 685--694, New York, NY, USA, 2011. ACM.

\bibitem[VV14]{valiant2014automatic}
Gregory Valiant and Paul Valiant.
\newblock An automatic inequality prover and instance optimal identity testing.
\newblock In {\em FOCS}, 2014.

\end{thebibliography}


\newcommand{\noopsort}[1]{} \newcommand{\printfirst}[2]{#1}
  \newcommand{\singleletter}[1]{#1} \newcommand{\switchargs}[2]{#2#1}
\begin{thebibliography}{10}
\providecommand{\url}[1]{#1}
\csname url@samestyle\endcsname
\providecommand{\newblock}{\relax}
\providecommand{\bibinfo}[2]{#2}
\providecommand{\BIBentrySTDinterwordspacing}{\spaceskip=0pt\relax}
\providecommand{\BIBentryALTinterwordstretchfactor}{4}
\providecommand{\BIBentryALTinterwordspacing}{\spaceskip=\fontdimen2\font plus
\BIBentryALTinterwordstretchfactor\fontdimen3\font minus
  \fontdimen4\font\relax}
\providecommand{\BIBforeignlanguage}[2]{{%
\expandafter\ifx\csname l@#1\endcsname\relax
\typeout{** WARNING: IEEEtran.bst: No hyphenation pattern has been}%
\typeout{** loaded for the language `#1'. Using the pattern for}%
\typeout{** the default language instead.}%
\else
\language=\csname l@#1\endcsname
\fi
#2}}
\providecommand{\BIBdecl}{\relax}
\BIBdecl

\bibitem{BatuKR04}
T.~Batu, R.~Kumar, and R.~Rubinfeld, ``Sublinear algorithms for testing
  monotone and unimodal distributions,'' in \emph{Proceedings of STOC}, 2004.

\bibitem{Bhattacharyya11}
A.~Bhattacharyya, E.~Fischer, R.~Rubinfeld, and P.~Valiant, ``Testing
  monotonicity of distributions over general partial orders,'' in \emph{ICS},
  2011, pp. 239--252.

\bibitem{batu2001testing}
T.~Batu, E.~Fischer, L.~Fortnow, R.~Kumar, R.~Rubinfeld, and P.~White,
  ``Testing random variables for independence and identity,'' in
  \emph{Proceedings of FOCS}, 2001.

\bibitem{alon2007testing}
N.~Alon, A.~Andoni, T.~Kaufman, K.~Matulef, R.~Rubinfeld, and N.~Xie, ``Testing
  k-wise and almost k-wise independence,'' in \emph{Proceedings of STOC}, 2007.

\bibitem{valiant2014automatic}
G.~Valiant and P.~Valiant, ``An automatic inequality prover and instance
  optimal identity testing,'' in \emph{FOCS}, 2014.

\bibitem{Birge87}
L.~Birg{\'e}, ``Estimating a density under order restrictions: Nonasymptotic
  minimax risk,'' \emph{The Annals of Statistics}, vol.~15, no.~3, pp.
  995--1012, September 1987.

\bibitem{hall2005testing}
P.~Hall and I.~Van~Keilegom, ``Testing for monotone increasing hazard rate,''
  \emph{Annals of Statistics}, pp. 1109--1137, 2005.

\bibitem{ChanDSS13b}
S.~O. Chan, I.~Diakonikolas, R.~A. Servedio, and X.~Sun, ``Efficient density
  estimation via piecewise polynomial approximation,'' in \emph{Proceedings of
  STOC}, 2014.

\bibitem{cule2010theoretical}
M.~Cule, R.~Samworth \emph{et~al.}, ``Theoretical properties of the log-concave
  maximum likelihood estimator of a multidimensional density,''
  \emph{Electronic Journal of Statistics}, vol.~4, pp. 254--270, 2010.

\bibitem{Paninski08}
L.~Paninski, ``A coincidence-based test for uniformity given very sparsely
  sampled discrete data.'' \emph{IEEE Transactions on Information Theory},
  vol.~54, no.~10, 2008.

\bibitem{AcharyaD15}
J.~Acharya and C.~Daskalakis, ``Testing poisson binomial distributions,'' in
  \emph{Proceedings of SODA 2015}.

\bibitem{CanonneDGR15}
C.~Canonne, I.~Diakonikolas, T.~Gouleakis, and R.~Rubinfeld, ``Testing shape
  restrictions,'' 2015, personal communication.

\bibitem{GibbsS02}
A.~L. Gibbs and F.~E. Su, ``On choosing and bounding probability metrics,''
  \emph{International Statistical Review}, vol.~70, no.~3, pp. 419--435, dec
  2002.

\bibitem{DvoretzkyKW56}
A.~Dvoretzky, J.~Kiefer, and J.~Wolfowitz, ``Asymptotic minimax character of
  the sample distribution function and of the classical multinomial
  estimator,'' \emph{The Annals of Mathematical Statistics}, vol.~27, no.~3,
  pp. 642--669, 09 1956.

\bibitem{Massart90}
P.~Massart, ``The tight constant in the {D}voretzky-{K}iefer-{W}olfowitz
  inequality,'' \emph{The Annals of Probability}, vol.~18, no.~3, pp.
  1269--1283, 07 1990.

\bibitem{AcharyaJOS14a}
J.~Acharya, A.~Jafarpour, A.~Orlitsky, and A.~T. Suresh, ``Efficient
  compression of monotone and $m$-modal distributions,'' in \emph{Proceedings
  of ISIT}, 2014.

\bibitem{KamathOPS15}
S.~Kamath, A.~Orlitsky, D.~Pichapati, and A.~T. Suresh, ``On learning
  distributions from their samples,'' in \emph{COLT}, 2015.

\bibitem{ChanDSS13a}
S.~O. Chan, I.~Diakonikolas, R.~A. Servedio, and X.~Sun, ``Learning mixtures of
  structured distributions over discrete domains,'' in \emph{Proceedings of
  SODA}, 2013.

\end{thebibliography}
